%% file: main.tex
\documentclass[fleqn,10pt]{wlscirep}

\usepackage{CJKutf8}

\usepackage{tabularx}
\usepackage[nolist]{acronym}
\usepackage{xargs} %
\usepackage{xspace}
\usepackage{multirow}
\usepackage{nicefrac}
\usepackage{float}
\usepackage{bbding}

\newcommand{\boxplotdef}{Standard box plot with Q1, median, Q3; whiskers at $1.5\times$ interquartile range.}
\usepackage{soul}

\title{Health-promoting Potential of Parks in 35 Cities Worldwide}

 \author[1,*]{Linus W. Dietz}
\author[2,3]{Sanja Šćepanović}
 \author[2,5]{Ke Zhou}
 \author[4]{André Felipe Zanella}
  \author[1,2,6]{Daniele Quercia}
 \affil[1]{King's College London, London, UK}
 \affil[2]{Nokia Bell Labs, Cambridge, UK}
   \affil[3]{University of Oxford, Oxford, UK}
 \affil[4]{Telefónica Innovación Digital, Barcelona, Spain}
   \affil[5]{University of Nottingham, Nottingham, UK}
  \affil[6]{Politecnico di Torino, Turin, Italy}
\affil[*]{linus.dietz@kcl.ac.uk}

\begin{abstract}

Urban parks are important for public health, but the role of specific spaces, such as playgrounds or lakes, and elements, such as benches or sports equipment, in supporting well-being is not well understood. Based on expert input and a review of the literature, we defined six types of health-related activities: physical, mindfulness, nature appreciation, environmental, social, and cultural. We built a lexicon that links each activity to specific elements and spaces within parks present in OpenStreetMap. Using this data, we scored 23,477 parks across 35 cities worldwide based on their ability to support these activities. We found clear patterns: parks in North America focus more on physical activity, while those in Europe offer more chances to enjoy nature. Parks near city centers support health-promoting activities better than those farther out. Suburban parks in many cities lack the spaces and equipment needed for nature-based, social, and cultural activities. We also found large gaps in park quality between cities. Tokyo and Paris provide more equal access, while Copenhagen and Rio de Janeiro show sharp contrasts. These results can help cities create fairer parks that better support public health.
\end{abstract}

\begin{document}

\begin{acronym}
\acro{LLM}{large language model}
\acro{OSM}{OpenStreetMap}
\acro{PCC}{Pearson Correlation Coefficent}
\acro{LBSN}{location-based social network}

\end{acronym}
\renewcommand\sectionautorefname{Section}
\renewcommand\subsectionautorefname{Section}
\renewcommand\subsubsectionautorefname{Section}
\flushbottom
\maketitle

{\def\thefootnote{}\footnotetext{Project website: \url{https://social-dynamics.net/healthy-parks}}
{\def\thefootnote{}\footnotetext{Replication Repository: \url{https://github.com/LinusDietz/Health-Promoting-Parks-Replication}}

\section{Main}
\input{sections/introduction.tex}

\section{Results}
\label{sec:results}
\input{sections/results.tex}

\section{Discussion}

\input{sections/discussion.tex}

\section{Methods}
\input{sections/methods.tex}

\section*{Code availability}
The Python and R code to compute park health scores is available at \\
\url{https://github.com/LinusDietz/Health-Promoting-Parks-Replication}.

\section*{Author contributions statement}

L.D. conceived and conducted the experiments and co-drafted the manuscript with S.S.
K.Z., A.F.Z. and D.Q. conceived the experiments and edited the manuscript.
All authors analyzed the results and reviewed the manuscript.

\section*{Ethics declarations}
\subsection*{Competing interests}
The authors declare no competing interests.

\clearpage

\input{sections/appendix.tex}

\end{document}

%% file: sections/introduction.tex
As the world's population continues to gravitate towards urban areas, cities are faced with the immense task of creating and maintaining green spaces to foster public health\cite{Kadakia2023Urbanization,Wu2023Improved}.
Urban parks are especially beneficial to vulnerable population groups such as socioeconomically deprived\cite{Maas2006Green} and older people\cite{Dzhambov2014Elderly,10.1136/jech.56.12.913}.
Yet the provision of amenities and facilities that support health-promoting activities has not been systematically analyzed at scale. %

Urban green spaces can support health in five main ways. First, they help people stay active, which can reduce obesity and heart diseases\cite{Hartig2014,twohig2018health}. Second, they give people a calm space to rest, reduce stress, and improve focus~\cite{Ulrich1984,ulrich1999effects,Hartig2014,Berman2008}. Third, they bring people together and support social ties~\cite{Jennings2019}. Fourth, they clean the air, block noise, and cool the city~\cite{Hartig2014,Nieuwenhuijsen2017}. Last, some green areas expose people to biodiverse forms of life, which may improve the immune system~\cite{Rook2013,Hanski2012}.

Park planners have often focused on practical ways to boost public health, such as adding sports fields, paths, or playgrounds~\cite{Barton2015Routledge,Gall2018Lessons,Kaczynski2008Association,Pietilae2015Relationships}.
Despite groving evidence of positive effects, spaces that allowed a diverse range of people to relax or enjoy beauty were often neglected in the past\cite{Campbell2001Rethinking}. Parks also have the power to build stronger communities\cite{campbell2016social}, and many planners now recognize that parks offer clear environmental gains such as cleaner air and lower noise levels\cite{Nieuwenhuijsen2017}. During the COVID-19 pandemic, cities saw how vital parks are for public well-being. This led to stronger calls for fair and flexible park design\cite{PATWARY2024124284}.

Most past studies have treated parks as simple green zones, based on size or distance from homes~\cite{Brown2008Theory,TU2020Travel}. Fewer have studied how park layout and equipment affect how people use them. For instance, SOPARC is a tool that tracks how people move through parks and what spaces they use~\cite{mckenzie2006soparc}. Other tools look at what makes parks good for activity, such as access, safety, or looks~\cite{bedimo2006development}. Still others focus on blue spaces such as rivers or lakes~\cite{mishra2020development}. These tools work well but are time-consuming to implement and hard to scale, and most focus only on physical activity~\cite{floyd2008park,kaczynski2014are}.
We still lack a clear understanding of the net benefits of parks, especially when weighing their positive aspects, such as providing spaces for recreation and supporting urban health\cite{Jimenez2021Associations,Maas2006Green}, with potential downsides, including reinforcing disparities \cite{Xiao2019Exploring,Wolch2014Urban,Bolte2025} or contributing to gentrification\cite{Anguelovski2022,Rigolon2020GreenGentrification}. 

For many cities, there is no easy way to list what parks contain or link these parts to health uses.
In this study, we score parks in 35 cities based on how well they support six types of health-related activity. We then ask how these scores relate to fair access to well equipped parks. Our work answers three questions. First, which health-related activities do different park spaces support? Second, how can we measure parks around the world based on this support? Third, where are the biggest gaps in what parks provide?

%% file: sections/results.tex
We began by creating a lexicon that linked park spaces and elements to health-related activities using \ac{OSM} data. We then scored parks in 35 major cities, which we selected for their geographic diversity and reliable \ac{OSM} coverage. After validating these scores through statistical comparison and expert review, we analyzed differences in park offerings within and between cities.

\subsection{Lexicon of Health-Related Park Features}

\subsubsection*{Step 1: Defining Activities in Parks}

An expert panel of three researchers with expertise in urban computing, Earth observation, and computational social science, identified common park activities and organized them into six categories:

\begin{description}[noitemsep,topsep=0pt,parsep=0pt,partopsep=0pt]
    \item \emph{Physical activities:} movement and sport such as walking, biking, swimming, and group exercise;
    \item \emph{Mindfulness activities:} practices like yoga, meditation, and tai chi;
    \item \emph{Nature appreciation:} observing and enjoying the natural environment, including bird watching and picnicking;
    \item \emph{Environmental activities:} community involvement such as gardening and conservation;
    \item \emph{Social activities:} gatherings and group events such as festivals and volunteering;
    \item \emph{Cultural activities:} heritage and arts programs, including performances and exhibitions.
\end{description}

While our 6 categories occasionally overlapped, they captured park activities at a more useful granular level than the typical recreational/physical division. Some overlap is inevitable since certain activities like football (with social and physical aspects) are inherently multi-faceted.

\subsubsection*{Step 2: Linking Facilities to Activities}

We used \ac{OSM} to collect elements and spaces located within parks in the 35 cities. These included defined areas like forests and ponds as well as features such as benches and sports courts. Each item was described with a tag, which we used to assign the item to an activity category. To scale this process, we trained a classifier based on an \ac{LLM} and validated it with an expert-coded dataset. The full lexicon included $1,441$ \ac{OSM} tags.  The ten most frequent tags for each category are listed in \autoref{tab:activities_osm_tags} in Appendix. No \ac{OSM} tags matched the mindfulness category, so we did not include it in our scoring. Future research may incorporate behavioral data to address this gap. %

\begin{figure}[t!]
	\includegraphics[width=\textwidth]{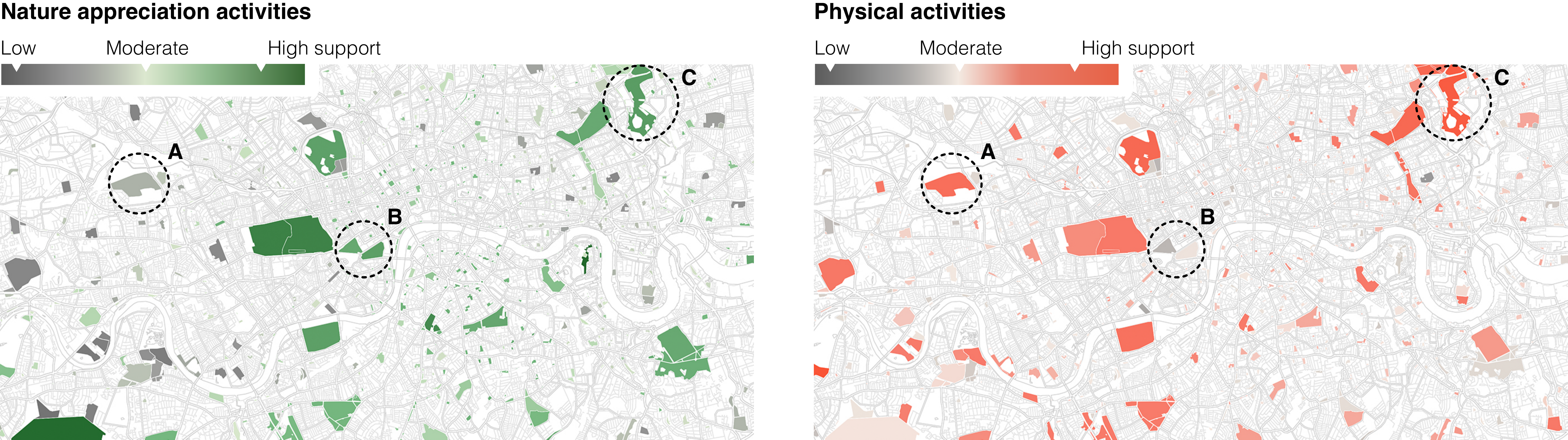}
	\caption{\textbf{Scores for nature appreciation (left) and physical activities (right) in London parks.} Some parks score low on nature appreciation but high on physical activity (A), while imperial-era parks such as Green Park and St. James's Park, close to the royal palace score high on nature appreciation but low on physical activity (B). Queen Elizabeth Olympic Park, purpose-built for the 2012 Games, scores high across all dimensions (C).\\ Map data from OpenStreetMap~\url{https://www.openstreetmap.org/copyright}}
	\label{fig:london_viz}
\end{figure}

\subsection{Scoring Parks for Health Support}
We scored each park based on the number of activity-related elements and spaces it contained, normalized by park area. Each park received a score for every activity category. \autoref{fig:london_viz} presents an example of these scores for London. 
In total, we scored 23,477 parks across 35 cities on five continents (\autoref{tab:correlation_flickr}). Each score reflects how well a park supports a specific activity compared to other parks of similar size in the same city.

\subsubsection*{Validation with Flickr Data}

\begin{table}[!t]
	\centering\footnotesize
	\caption{\textbf{Our activity scores correlate with the prevalence of photographed activities from Flickr images.} Pearson Correlation Coefficients (PCC) between \ac{OSM} scores and Flickr activity scores for parks with at least 250 photos.}
	\label{tab:correlation_flickr}
	\input{tables/city_flickr_correlations.tex}
\end{table}

We validated our park scores using geotagged Flickr images taken within park boundaries. These images included both user-generated and machine-generated tags. We used sentence embeddings to match these tags to our activity taxonomy, allowing us to estimate activity scores for each park based on photo content.

We compared these Flickr-based activity scores with our \ac{OSM}-derived scores using the \ac{PCC}. Across the 35 cities, the average correlation was $0.39$ with a standard deviation of $0.07$, indicating moderate but consistent agreement between the two sources. The highest agreement appeared in the cultural ($\mu = 0.53$), social ($\mu = 0.39$), and environmental ($\mu = 0.39$) categories. These types of activities often involve distinctive and photogenic features such as festivals, artworks, or gardens, which are likely to be photographed and tagged.

In contrast, the physical ($\mu = 0.30$) and nature-appreciation ($\mu = 0.33$) categories showed weaker correlations. We believe this may be because users often photograph people or scenery without tagging specific sports or nature-related elements, making it harder to detect those activities in the data.

At the city level, Washington, DC ($\mu = 0.52$) and Perth ($\mu = 0.50$) had the strongest correlations. These cities are in English-speaking countries where Flickr and \ac{OSM} usage is more common, which likely led to higher data quality and tag relevance. Cities with lower correlations, including Amsterdam, Hong Kong, and Vienna (each $\mu \cong 0.3$), may have had fewer geotagged images, lower tag accuracy, or less alignment with our English-based taxonomy.

These findings support the use of \ac{OSM} data to assess park infrastructure for health-promoting activities, especially in categories and regions where online content about park use is widely available.

\subsubsection*{Validation with Wikipedia}

We also validated our activity scores using Wikipedia pages. We identified the top-scoring parks globally for each health-related activity based on our scoring system. For each park, we examined the Wikipedia page to confirm whether the park was known for the corresponding activity type.

In the \emph{physical} category, we found that parks such as Centennial Parklands in Australia and Bois de Boulogne in Paris were known for extensive sports infrastructure, including fields for tennis, soccer, and polo. For \emph{nature appreciation}, top parks like La Dehesa de la Villa in Madrid and Guandu Nature Park in Taipei stood out in their cities with large green areas and wildlife. In the \emph{environmental} category, Washington Park Arboretum in Seattle and Kita-no-maru Park in Tokyo included botanical gardens and conservation efforts. \emph{Social} activity leaders like Inspiration Lake in Hong Kong and Toronto Island Park in Canada were cited for hosting gatherings and public amenities. In the \emph{cultural} category, Ueno Park in Tokyo featured museums and historic landmarks, while Seattle Center included concert halls and art venues.

To further verify our assignments, we conducted structured search queries, such as ``Tokyo parks for cultural activities'' and found that the top-ranked parks by our method consistently appeared in online recommendations and guides. A complete list of these parks, organized by city and activity category, is available in \autoref{tab:top_city_parks}.

\begin{figure}[t]
	\includegraphics[width=\textwidth]{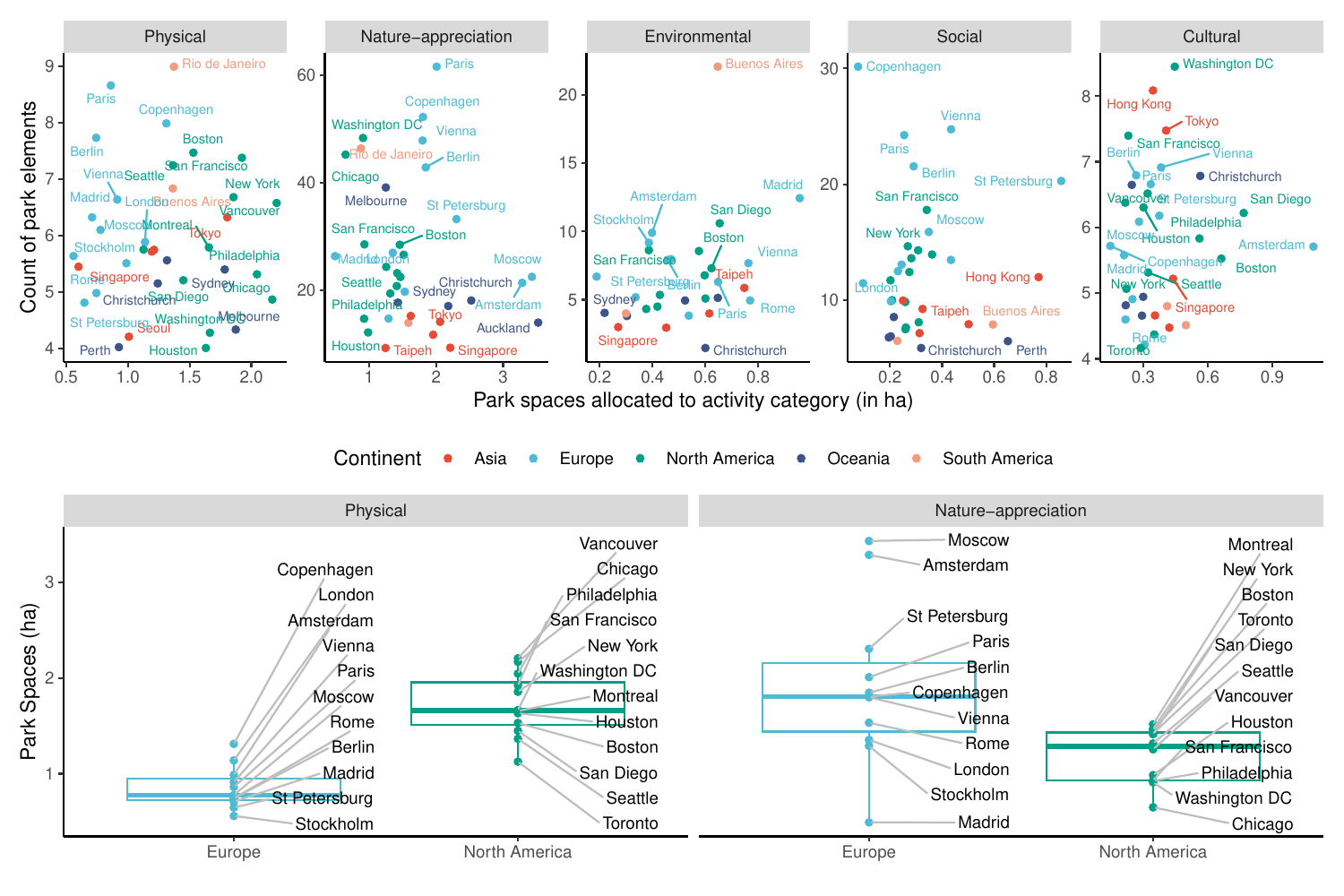}
	\caption{\textbf{Top: Area of park spaces (horizontal axis) and count of elements (vertical axis) in a typical 8-hectare park by activity and city.} Nature and physical activity spaces are the most frequent. Nature-related elements are also the most common across cities. Cultural spaces and elements appear least often. Axes are scaled for clarity.\\
	\textbf{Bottom: Average space in 8-hectare parks dedicated to nature and physical activities in North America and Europe.} North American parks support physical activity more, while European parks provide more space for nature appreciation. \boxplotdef}
	\label{fig:8ha_park_counts}
\end{figure}

\subsection{Cities Differ in Health-Related Activity Support}
Our analysis revealed systematic variations in how urban parks support health-promoting activities, with differences occurring both between cities and within individual urban areas. These variations reflect distinct urban planning priorities, geographic influences, and equity considerations that shape park design and resource allocation.

\subsubsection*{Estimating Park Offerings Through 8-Hectare Model Parks}

To compare park features systematically across cities, we constructed statistical models of typical 8-hectare parks: a size representing neighborhood parks and matching our dataset's average. \autoref{fig:8ha_park_counts} (top) shows the expected number of elements and area dedicated to each activity type in a \emph{statistically average} 8-hectare park for each city. We computed these values using parameters from linear regression models developed for each city (\autoref{eq:linear_log_models}).  While we fitted separate models per city, making raw health scores not directly comparable between cities, the models enabled comparison of hypothetical average parks across urban contexts.

\subsubsection*{Regional Priorities: European Nature Focus vs. North American Physical Activity}

The most notable regional contrast lies in emphasis on physical activity versus nature appreciation, as shown in \autoref{fig:8ha_park_counts}~(bottom). Parks in European cities tended to provide more space for nature appreciation, while North American cities allocated more area to physical activities. These differences reflect broader urban planning patterns and suggest that parks serve different roles depending on regional goals and cultural expectations.

\subsubsection*{Cultural Features: Universally Limited Support}

Across all regions, cultural features appeared least frequently in parks. In terms of elements, nature appreciation appeared most often, followed by social and environmental activities. Cultural features were the least common. This pattern suggests that while nature and movement receive widespread support, cultural health-promoting activities are deprioritized globally in urban park design.

\subsection{Park Offerings are Unequally Distributed Within Cities}

\label{sec:geographic_influence}

\begin{figure}[t]
 \centering
\includegraphics[width=0.88\textwidth]{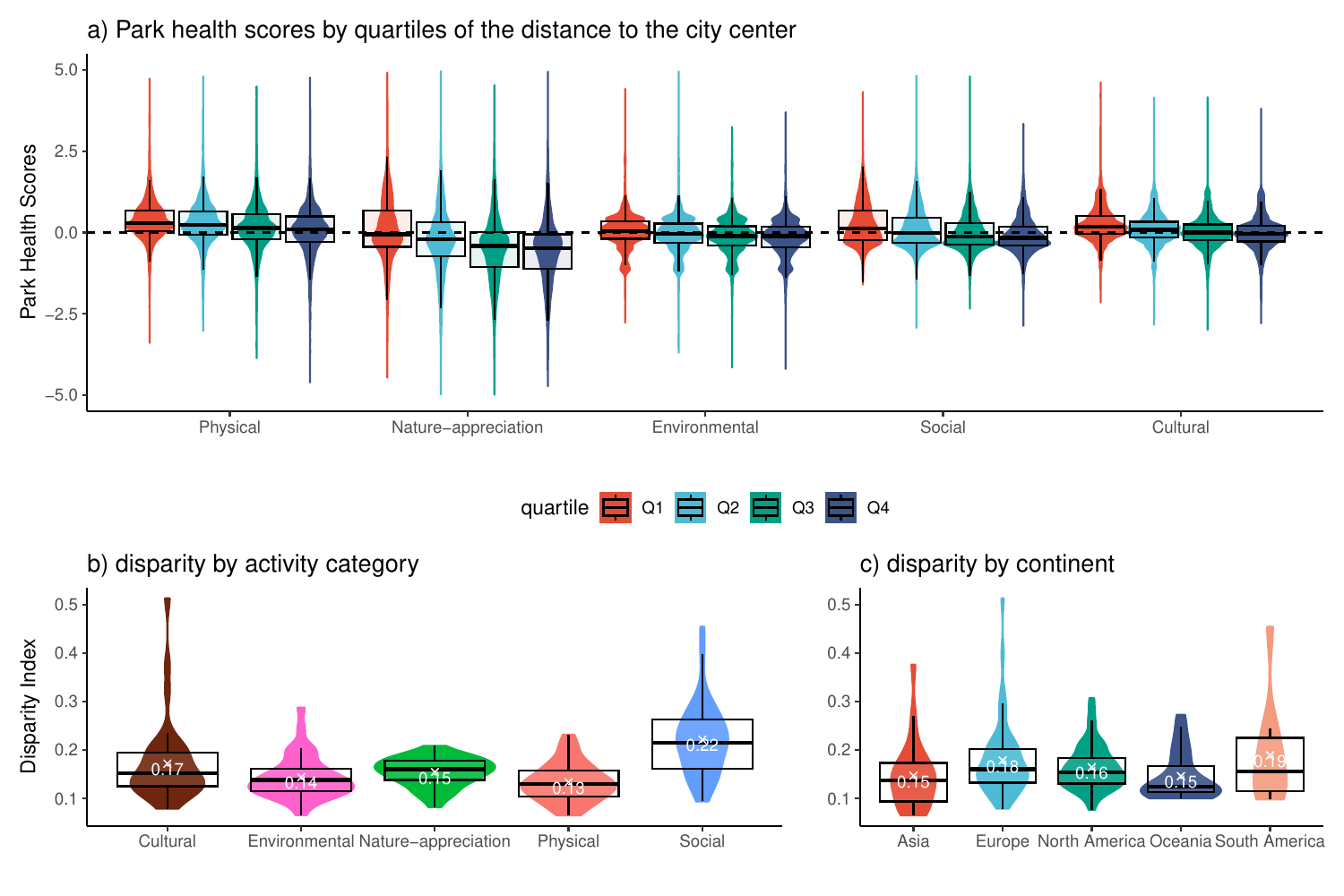}
	\caption{\textbf{(a) Park scores by distance quartile.} Q1 includes parks closest to the city center; Q4 includes those farthest away. Scores decline steadily with distance, especially for nature appreciation. Environmental features show the least decline. \boxplotdef~
	\textbf{(b, c) Inequality in park health scores.} Disparities across activities and continents are shown. While total disparity ranges are similar across regions, some activities, such as social and cultural, show much higher inequality than others.}
	\label{fig:quartiles_scores}
\end{figure}

We examined how park scores vary within cities by grouping parks into quartiles based on distance from city centers, from the innermost 25\% (Q1) to the outermost 25\% (Q4). This approach allowed us to systematically assess whether proximity to urban centers influences park offerings.

\subsubsection*{Inner-City Parks Consistently Outperform Peripheral Ones}

Across all activity types, parks in central areas scored higher. \autoref{fig:quartiles_scores}a demonstrates that park scores dropped steadily with each quartile further from the center. This pattern held across most cities on all continents, with exceptions in Buenos Aires and Rio de Janeiro, where geography and coastal form influence green space layout.

\subsubsection*{Nature Appreciation Shows Strongest Distance-Related Decline}

The trend was most pronounced for nature appreciation. This finding challenges assumptions that outer areas naturally support greener, more restorative environments and echo Montgomery's concept of the ``Savannah Trap\cite{Montgomery2015happy}'', where suburban areas include empty, open land that does not support social or ecological use.

\subsection{Cities Vary in Ensuring Equitable Park Access}
\label{sec:results_inequality_continents}

\begin{table}[!t]
	\centering\footnotesize
	\caption{\textbf{The disparity scores by city and activity type are generally low, but vary based on the activity category.} Cities are ordered by their overall average disparity. Lower values indicate a more even distribution of features. Scores one standard deviation below or above the mean are marked in gray and bold, respectively.}
	\label{tab:gini_cities}
	\input{tables/gini_cities.tex}
\end{table}

We assessed equity in park offerings within cities using a disparity index (\autoref{eq:gini_scores}) ranging from 0~(complete equality) to 1~(maximum inequality). Box plots in \autoref{fig:quartiles_scores}b–c show average disparities by activity and region, with detailed values in \autoref{tab:gini_cities}.

\subsubsection*{Activity-Specific Disparity Patterns}

Disparities varied by activity type: \emph{physical activities} showed the most even distribution (\(\mu=0.135, \sigma=0.039\)), while \emph{social activities} exhibited the highest disparity (\(\mu=0.222, \sigma=0.073\)), indicating that some parks offer rich social features while others lack them entirely. \emph{Nature appreciation} had moderate average disparity (\(\mu=0.157\)) but low variation (\(\sigma=0.032\)), suggesting consistent global trends. Overall disparities averaged $0.165$, and were similar across continents ($0.15$--$0.19$), suggesting that inequality in park design is not limited to one region.

\subsubsection*{Cities Achieving Balanced Activity Support}
\label{sec:results_inequality_gini}

Tokyo, Paris, Auckland, Buenos Aires, and Hong Kong demonstrated the most equitable park offerings across activities. Tokyo and Paris particularly excelled, with four or more activity categories showing disparities below one standard deviation from the mean. These results may reflect centralized park governance, strong public policy, and active community involvement in cities such as Paris \cite{Jole2008public} and Tokyo \cite{fujita2016residential}.

\subsubsection*{Cities with Concentrated Park Resources}

In contrast, Moscow, St. Petersburg, Stockholm, Rio de Janeiro, and Copenhagen showed high disparities in park offerings across categories. Stockholm exhibited the highest single-category disparity in cultural features ($0.514$), suggesting cultural resources were concentrated in a few parks. Copenhagen showed high disparity in both cultural and social categories, potentially reflecting gentrification\cite{copenhagen_ineq}. Rio de Janeiro displayed widespread disparities across social, environmental, and cultural categories, aligning with prior findings on spatial health inequality\cite{bortz2015disaggregating}.

These findings demonstrate that while park inequality exists globally, specific cities have successfully implemented more equitable approaches to supporting diverse health-promoting activities through their urban park systems.

%% file: tables/city_flickr_correlations.tex
\begin{tabular}{lrrrrrrr}
  \toprule
\multirow{2}{*}{\bfseries City} & \multirow{2}{*}{\bfseries Parks} & \multirow{2}{*}{\bfseries Mean PCC} & \multicolumn{5}{c}{\bfseries Individual Activity Categories PCC} \\ \cmidrule(lr){4-8} & & & \textbf{Physical} &  \textbf{Nature-appreciation}  & \textbf{Environmental} & \textbf{Social} & \textbf{Cultural}  \\ 
  \midrule
Amsterdam & 25 & 0.32 & 0.24 & 0.36 & 0.2 & 0.49 & 0.29 \\ 
  Auckland & 33 & 0.46 & 0.38 & 0.39 & 0.43 & 0.47 & 0.63 \\ 
  Berlin & 77 & 0.34 & 0.22 & 0.14 & 0.53 & 0.39 & 0.42 \\ 
  Boston & 50 & 0.4 & 0.27 & 0.43 & 0.36 & 0.35 & 0.58 \\ 
  Buenos Aires & 54 & 0.36 & -0.04 & 0.34 & 0.92 & 0.15 & 0.41 \\ 
  Chicago & 69 & 0.41 & 0.2 & 0.45 & 0.35 & 0.45 & 0.6 \\ 
  Christchurch & 16 & 0.49 & 0.27 & 0.52 & 0.21 & 0.72 & 0.71 \\ 
  Copenhagen & 19 & 0.33 & 0.31 & 0.31 & 0.01 & 0.22 & 0.8 \\ 
  Hong Kong & 80 & 0.3 & 0.38 & 0.21 & 0.36 & 0.21 & 0.32 \\ 
  Houston & 28 & 0.46 & 0.36 & 0.48 & 0.15 & 0.61 & 0.68 \\ 
  London & 304 & 0.44 & 0.45 & 0.4 & 0.32 & 0.45 & 0.55 \\ 
  Madrid & 42 & 0.32 & 0.08 & -0.04 & 0.19 & 0.66 & 0.72 \\ 
  Melbourne & 59 & 0.49 & 0.45 & 0.63 & 0.44 & 0.64 & 0.29 \\ 
  Montreal & 55 & 0.43 & 0.31 & 0.27 & 0.44 & 0.49 & 0.64 \\ 
  Moscow & 65 & 0.34 & 0.33 & 0.2 & 0.34 & 0.21 & 0.6 \\ 
  New York & 210 & 0.42 & 0.41 & 0.21 & 0.58 & 0.45 & 0.47 \\ 
  Paris & 108 & 0.39 & 0.42 & 0.28 & 0.3 & 0.39 & 0.55 \\ 
  Perth & 23 & 0.5 & 0.18 & 0.71 & 0.3 & 0.68 & 0.61 \\ 
  Philadelphia & 39 & 0.46 & 0.43 & 0.2 & 0.63 & 0.32 & 0.75 \\ 
  Rio de Janeiro & 19 & 0.36 & 0.26 & 0.6 & 0.23 & 0.41 & 0.33 \\ 
  Rome & 41 & 0.31 & 0.43 & 0.35 & 0.08 & 0.16 & 0.52 \\ 
  San Diego & 47 & 0.44 & 0.27 & 0.38 & 0.61 & 0.48 & 0.44 \\ 
  San Francisco & 98 & 0.31 & 0.17 & 0.19 & 0.42 & 0.32 & 0.46 \\ 
  Seattle & 76 & 0.38 & 0.47 & 0.36 & 0.09 & 0.4 & 0.6 \\ 
  Seoul & 52 & 0.35 & 0.34 & 0.36 & 0.43 & 0.23 & 0.38 \\ 
  Singapore & 73 & 0.36 & 0.14 & 0.07 & 0.61 & 0.39 & 0.6 \\ 
  St Petersburg & 28 & 0.34 & 0.17 & 0.22 & 0.38 & 0.39 & 0.56 \\ 
  Stockholm & 50 & 0.47 & 0.45 & 0.32 & 0.93 & 0.23 & 0.42 \\ 
  Sydney & 99 & 0.31 & 0.42 & 0.39 & 0.22 & 0.14 & 0.36 \\ 
  Taipeh & 107 & 0.34 & 0.16 & 0.22 & 0.48 & 0.26 & 0.56 \\ 
  Tokyo & 208 & 0.31 & 0.19 & 0.24 & 0.37 & 0.32 & 0.42 \\ 
  Toronto & 111 & 0.34 & 0.31 & 0.3 & 0.17 & 0.39 & 0.51 \\ 
  Vancouver & 62 & 0.44 & 0.33 & 0.37 & 0.33 & 0.5 & 0.67 \\ 
  Vienna & 40 & 0.3 & 0.18 & 0.08 & 0.57 & 0.19 & 0.5 \\ 
  Washington DC & 61 & 0.52 & 0.4 & 0.52 & 0.56 & 0.56 & 0.56 \\ 
  \midrule
  Mean (sd) & 72.23 (60.07) & 0.39 (0.07) & 0.29 (0.12) & 0.33 (0.16) & 0.39 (0.21) & 0.39 (0.16) & 0.53 (0.13) \\ 
   \bottomrule
\end{tabular}

%% file: tables/gini_cities.tex
\begin{tabular}{lrrrrrr}
  \toprule
\multirow{2}{*}{\bfseries City} & \multirow{2}{*}{\bfseries Mean Score} & \multicolumn{5}{c}{\bfseries Inequality of each activity category's offering in a city.} \\ \cmidrule(lr){3-7} &  & \textbf{Physical} & \textbf{Nature-appreciation} &  \textbf{Environmental} & \textbf{Social} & \textbf{Cultural}  \\ 
  \midrule
Tokyo & {\color{gray}0.089} & {\color{gray}0.065} & {\color{gray}0.082} & {\color{gray}0.066} & {\color{gray}0.146} & {\color{gray}0.084} \\ 
  Paris & {\color{gray}0.097} & 0.104 & {\color{gray}0.101} & {\color{gray}0.087} & {\color{gray}0.114} & {\color{gray}0.079} \\ 
  Auckland & {\color{gray}0.121} & 0.103 & {\color{gray}0.117} & 0.105 & 0.154 & 0.126 \\ 
  Buenos Aires & {\color{gray}0.125} & 0.172 & {\color{gray}0.107} & {\color{gray}0.098} & {\color{gray}0.137} & 0.108 \\ 
  Hong Kong & {\color{gray}0.125} & {\color{gray}0.079} & 0.134 & \bfseries{0.197} & {\color{gray}0.094} & 0.12 \\ 
  Vienna & 0.135 & 0.13 & 0.144 & 0.107 & 0.17 & 0.125 \\ 
  New York & 0.138 & 0.113 & 0.137 & 0.131 & 0.177 & 0.129 \\ 
  Rome & 0.138 & 0.122 & 0.16 & 0.116 & 0.153 & 0.141 \\ 
  Christchurch & 0.138 & 0.104 & {\color{gray}0.113} & 0.101 & 0.273 & 0.101 \\ 
  San Francisco & 0.14 & {\color{gray}0.09} & 0.152 & 0.116 & 0.212 & 0.132 \\ 
  Berlin & 0.143 & 0.131 & 0.14 & 0.138 & 0.152 & 0.154 \\ 
  Chicago & 0.153 & {\color{gray}0.077} & 0.178 & \bfseries{0.205} & 0.151 & 0.153 \\ 
  San Diego & 0.153 & 0.129 & 0.169 & 0.124 & 0.214 & 0.128 \\ 
  Melbourne & 0.155 & 0.122 & 0.167 & 0.124 & 0.248 & 0.114 \\ 
  Montreal & 0.156 & 0.126 & 0.178 & 0.142 & 0.215 & 0.118 \\ 
  Singapore & 0.157 & 0.166 & 0.14 & 0.119 & 0.227 & 0.133 \\ 
  Perth & 0.157 & 0.132 & \bfseries{0.19} & 0.125 & 0.21 & 0.128 \\ 
  Madrid & 0.16 & 0.104 & 0.163 & 0.168 & 0.197 & 0.167 \\ 
  Sydney & 0.16 & 0.11 & {\color{gray}0.119} & 0.153 & 0.226 & 0.191 \\ 
  London & 0.163 & 0.134 & 0.154 & 0.138 & 0.23 & 0.158 \\ 
  Vancouver & 0.165 & 0.102 & 0.144 & 0.153 & 0.216 & 0.208 \\ 
  Toronto & 0.169 & 0.129 & 0.178 & 0.124 & 0.24 & 0.172 \\ 
  Houston & 0.173 & 0.139 & \bfseries{0.188} & 0.141 & 0.218 & 0.18 \\ 
  Taipeh & 0.174 & {\color{gray}0.082} & 0.156 & 0.116 & {\color{gray}0.138} & \bfseries{0.377} \\ 
  Philadelphia & 0.175 & 0.158 & 0.174 & 0.122 & \bfseries{0.308} & 0.116 \\ 
  Seattle & 0.179 & 0.132 & 0.161 & 0.141 & 0.265 & 0.195 \\ 
  Boston & 0.186 & 0.158 & 0.183 & 0.147 & 0.292 & 0.152 \\ 
  Washington DC & 0.191 & 0.154 & 0.16 & 0.173 & 0.262 & 0.205 \\ 
  Seoul & 0.191 & \bfseries{0.177} & 0.184 & 0.174 & 0.271 & 0.152 \\ 
  Amsterdam & 0.197 & 0.161 & \bfseries{0.187} & 0.153 & 0.267 & 0.217 \\ 
  Moscow & \bfseries{0.205} & \bfseries{0.232} & 0.161 & \bfseries{0.205} & 0.236 & 0.194 \\ 
  St Petersburg & \bfseries{0.213} & \bfseries{0.21} & 0.166 & \bfseries{0.288} & 0.201 & 0.199 \\ 
  Stockholm & \bfseries{0.247} & \bfseries{0.189} & 0.132 & 0.105 & \bfseries{0.297} & \bfseries{0.514} \\ 
  Rio De Janeiro & \bfseries{0.254} & 0.141 & \bfseries{0.193} & \bfseries{0.245} & \bfseries{0.455} & 0.236 \\ 
  Copenhagen & \bfseries{0.266} & \bfseries{0.209} & \bfseries{0.209} & \bfseries{0.195} & \bfseries{0.398} & \bfseries{0.321} \\ 
  \midrule Mean (sd) & 0.165 (0.039) & 0.134 (0.039) & 0.155 (0.029) & 0.144 (0.045) & 0.222 (0.074) & 0.172 (0.085) \\ 
   \bottomrule
\end{tabular}

%% file: sections/discussion.tex
Urban parks offer health benefits that go beyond the usual physical and mental wellbeing. However, our global analysis reveals that urban parks require strategic improvement focused on geographic equity and activity diversity to maximize their health benefits.

\subsection{Main Findings}
By identifying six types of activities people do in parks, we created a detailed taxonomy of park activities and used this to evaluate parks worldwide based on their support for five of the six activity categories (data to evaluate the mindfulness category was not available on \ac{OSM}). By giving individual scores for each activity, we identified three critical areas that demand attention from landscape architects, urban designers, and policymakers.

\subsubsection*{Geographic Inequities Demand Immediate Action}

Our findings show a clear and concerning geographic trend: parks in city centers are better equipped for health-promoting activities than those on the outskirts. This pattern holds across all types of activities, including those that might be expected to be more common in less populated areas, such as nature appreciation or environmental activities. This geographic disparity contradicts basic expectations and creates significant barriers to equitable health access.

This finding aligns with previous research. Wolch et al. found that park distribution often favors more wealthy communities\cite{Wolch2014Urban}. Other studies have shown that recent urban development policies have led to significant greening in city centers, while suburban areas have received less investment\cite{Sun2020Dramatic}. Our results expand on these works by showing not only reduced access to parks in suburban areas\cite{Sun2020Dramatic,Wolch2014Urban}, but also a lack of amenities and spaces for health-promoting activities.

This evidence creates a clear call to action for urban planners to shift their focus to improving suburban parks, as well as parks outside urban centers, where significant and often overlooked recreation gaps are often found~\cite{kleinhewett2024classifying}. However, creating high-quality urban green spaces can be a balancing act, with potential downsides such as gentrification\cite{Rigolon2020GreenGentrification,Wolch2014Urban}. Designers must think carefully about the urban context when improving parks in ways that may change their role\cite{Rigolon2020GreenGentrification}.

\subsubsection*{Activity-Type Inequities Show Mixed but Encouraging Patterns}

We also examined disparities in the health potential of parks among cities by activity type. The findings are generally encouraging: overall disparity scores were low and did not show pronounced global variation, in contrast to earlier studies\cite{Chen2022Contrasting,Long2022Visualizing}. While previous work has shown high inequality in absolute access to greenery, our results suggest that relative inequality in access to health-promoting park activities is less severe.

The distribution patterns vary significantly by activity type, revealing both strengths and critical gaps. Physical activity infrastructure was the most evenly distributed, likely reflecting its prominence in public health discourse and urban policy\cite{Pietilae2015Relationships}. This aligns with the medical literature's emphasis on physical activity benefits (\autoref{tab:activities_literature_full}). In contrast, facilities for social activities were the most unequally distributed, revealing a key area for improvement\cite{Zhang2022Assessing}.

The global nature of this challenge is evident from our data: the five cities with the greatest disparities in park offerings span four continents, indicating that uneven provision of well-equipped parks is not a concentrated issue. This suggests that just urban planning is not only a matter of financial resources but also one of municipal priorities and community engagement\cite{Jole2008public,Jones2002Enticement}. A positive trend can be observed in the Asian countries of our study and Oceania, which have the most fair provision of park facilities.

\subsubsection*{Our Findings Provide Actionable Insights for the Design of Well-equipped Parks}

To understand the relevance of our work for urban planning, we conducted 30-minute semi-structured interviews with three domain experts (see \autoref{sec:expert_rubric}). We reached out to practitioners with 12--30 years of professional experience from diverse backgrounds: a lecturer and urban designer (E1, male, Hong Kong), an urban designer and master planner (E2, female, North America), and a municipal park development manager (E3, male, United Kingdom). The experts confirmed that our quantitative approach offers significant practical value for urban design decisions. They identified several ways our findings could support urban design: reducing reliance on subjective park assessments (E1), serving as a quantitative baseline for master planning (E1 and E2), and providing an evidence-based alternative to institutional knowledge and practice-specific heuristics (E1). The urban designers emphasized that the visualization of park scores is already useful to \emph{``identify the gaps regarding what is present in terms of offerings in a district (E1)''}, especially when aiming to ensure fair access to parks. This was echoed by the park manager, who stated they would use the park scoring to guide development priorities, and as supporting evidence to gain leadership backing or external funding. When shown the taxonomy of activities, they noted that most of their projects addressed physical and nature-appreciation activities, and \emph{``[the activity taxonomy] helps us think about what it is we're doing and how and how some categories have been neglected (E3).''}

The experts also highlighted the importance of the way our scores are normalized to the specific context of each city. Our park scores are normalized to reflect the unique context of each city, making direct comparisons between cities hard. While this limits inter-city comparisons, it avoids drawing misleading conclusions across culturally, geographically, and climatically diverse settings. For example, E1 cited culturally different approaches to park provision in the form of ``three-dimensional parks'' in high-density environments like Hong Kong, while E2 brought up the challenges in adopting urban transformation manuals from Western cities in regions with extreme climates such as the Middle East.

\subsection{Limitations and Future Directions}

Despite the well-known benefits of parks, lack of fair access to urban green spaces remains a common theme in academic literature\cite{Wolch2014Urban,Sun2020Dramatic,Chen2022Contrasting,Wu2023Improved,Kleinschroth2024Global}. We focused only on urban parks, excluding other types of urban green spaces such as gardens, street trees, and green roofs, which also contribute to urban health. Parks were selected because they support a wider range of activities than more specialized spaces, are generally publicly accessible, and are typically maintained by municipal administrations, making any identified shortcomings more actionable for policy and planning.

Relying on map data limits our analysis to activities that can be linked to specific elements and spaces within parks. While some studies explored the relationship between open spaces and their use\cite{Golicnik2010Emerging}, we avoided making assumptions about the usage of informal spaces to reduce the risk of cultural bias in our park evaluations. As a result, we excluded mindfulness activities from our main analysis, as they often do not require designated areas in city parks. Likewise, our data does not account for temporary cultural events, such as music festivals held in parks.

Park offerings were scored under the assumption that the presence of facilities and spaces enabling certain activities is a necessary condition for realizing specific health benefits. However, their presence alone may not be enough to deliver these benefits in practice. Furthermore, not all activities are directly tied to specific infrastructure, and \ac{OSM} cartographic data alone cannot capture the intensity of use. We did not distinguish between the quality of maintenance or design of individual elements, such as benches being equipped with backrests, or the size, layout and species of trees and other plants.

Like all urban spaces, parks are in constant flux, which raises the question of how parks have evolved over time~\cite{Wu2023Improved}. This became clear during the COVID-19 pandemic, when many cities had to rethink their strategies, possibly leading to big changes in park facilities with major social justice effects~\cite{10.3389/frsc.2021.710243}.
Looking at historical data from \ac{OSM} is challenging, as it is difficult to distinguish between actual changes in park facilities and the increasing completeness of the \ac{OSM} database.

A key direction for future research is to examine how park offerings directly influence health outcomes. While collecting global medical data to evaluate the health impact of individual parks is impractical, more targeted investigations may be feasible in specific contexts. For instance, prescription data available in some countries could be used to estimate the prevalence of certain medical conditions\cite{Scepanovic2024MedSat}, and to explore potential causal links between changes in park infrastructure and improvements in public health over time.

We showed that the \ac{OSM} park scores are closely linked to activities captured in Flickr photos, but other data could also be used to approximate park usage. For instance, detailed mobile traffic data could offer interesting insights into how parks are used\cite{Heikinheimo2020Understanding,Xiao2019Exploring,Zanella2025Digital}.
This could also help reduce biases inherent in social media data sources like Flickr and \ac{OSM}, which rely on actively submitted contributions from tech-savvy users, and sometimes out-of-city visitors rather than passive observations of citizen behavior.

Finally, by focusing on health-promoting activities, our work introduces avenues for health-related behavioral change\cite{Zhang2017conceptual,campbell2016social}. For instance, our activity-based approach could support context-aware recommendations\cite{Dietz2024Exploratory}, such as identifying the most suitable times to engage in specific park activities when conditions like temperature, air quality, or crowdedness are most favorable.

%% file: sections/methods.tex
\subsection{Data}
\label{subsec:data}

OSM is a globally encompassing geographic information database based on crowdsourced contributions.
While accessible through a map interface at \url{https://openstreetmap.org}, its primary value lies in being an indispensable source for open mapping data both in commercial and scientific applications~\cite{Li2022OpenStreetMap,Herfort2021evolution,Jacobs2020OpenStreetMap}.
Thanks to its permissive licensing, \ac{OSM} has fostered a large ecosystem of individual and professional contributors. 
As a result, the project has achieved comprehensive worldwide coverage, with near-perfect mapping quality across the western world\cite{Jacobs2020OpenStreetMap}, while retaining remarkable detail in the global south\cite{Herfort2021evolution}.
In this study, we utilized \ac{OSM} data from fall 2023.

\ac{OSM} employs a tagging system with key-value pairs to categorize and describe all these map objects.
Each map object is typically associated with multiple tags that describe its purpose, but may also include additional information, such as opening hours, or data source references.
To avoid the complexities of the \ac{OSM} data model, %
for our tasks, it was sufficient to focus on two key map objects related to parks: \emph{park elements} and \emph{park spaces}.
\emph{Park elements} are 0-dimensional points representing objects like benches, individual trees, and statues.
On the other hand, \emph{park spaces} refer to areas within the parks, such as meadows, lakes, and forests.

\vspace{0.05in}
\noindent \textbf{Flickr} (\url{https://flickr.com}) has established itself as one of the most prominent platforms for sharing photography.
Since its inception in 2004, the platform has gained considerable popularity, accumulating billions of images.
Notably, many of these images have been precisely geo-located, thanks to the utilization of the (phone) camera's GPS module.

We utilized a substantial dataset comprising geo-located images posted between 2004 and 2015. 
This extensive dataset offered us a valuable secondary perspective on activities taking place within the parks of the world.
By intersecting these images with the park outlines from \ac{OSM}, we identified 10,788,686 pictures captured within the boundaries of parks in our study cities.
To extract the depicted content from these images, we used user-assigned tags in conjunction with automatically-generated computer vision labels~\cite{Li2009Towards,Thomee2016YFCC100M}.

\subsection{Study Area}

Our research focused on 35 cities listed in \autoref{tab:correlation_flickr}, which we selected using three criteria to make our analysis broad yet robust.

First, we selected major cities worldwide with populations of at least 650,000. This threshold includes many of the largest urban areas, such as major European capitals and other densely populated regions where parks play a vital role in public well-being. To improve representation in Oceania, however, we made an exception for Christchurch, New Zealand, which has a smaller population. This first criterion allowed us to examine parks in cities from various parts of the world, each affected by its own climate, history, and cultural background.

Second, we only looked at cities in countries where at least 80\% of the population has access to the Internet~\cite{InternetUsers}. This ensured we had enough online data (like tags on OSM or photos on Flickr) for our study. Since there is no detailed global data on Internet use in cities specifically, we used the country's overall access to the Internet as our guide. %
We decided on this threshold upon our preliminary analyses, finding that in many cities in Africa and South America, there was not enough digital information for our approach, which relies on social media and collaborative mapping data.

Third, we chose cities where, on average, parks have at least one-eighth of park areas are annotated with health-related tags on \ac{OSM}. 
Since our analysis relied heavily on \ac{OSM} data, this criterion ensured a minimum level of information on the platform for our study. We settled on this one-eighth threshold after observing that, below this level, the lack of contributor-added tags limited our ability to extract meaningful information.
This primarily excluded cities where most tagging was done predominantly automatically through earth observation that was not accompanied by manual tagging of OSM contributors.
This was mainly the case in China, where non-governmental mapping is restricted\cite{Wen2018Volunteered}.

\subsection{Identifying Health-promoting Activities in Urban Greenery} 
\label{sec:literature_method}

We identified and categorized park-based activities using input from an expert panel consisting of three co-authors of this study. We compiled a comprehensive list of activities commonly undertaken in urban parks.
To collect relevant papers, we used two specific queries of Google Scholar: ``(urban) AND  (parks OR greenery) AND usage'' and ``(activities in urban) AND (parks OR greenery)''.
From this process, we retrieved the top 50 scholarly articles for each search phrase, resulting in a total of 91 unique papers.
We reviewed each article and collected all activities, resulting in a diverse set of activity descriptions varying in granularity. For example, the literature included both broad terms such as \emph{leisure activities} or \emph{recreation}, as well as more specific categories like \emph{physical} and \emph{social} activities. We also noted plenty of individual activities like \emph{walking}, \emph{performing street theater}, \emph{fishing}, and \emph{playing all kinds of different sports}.
Subsequently, we convened to categorize the identified activities, with a particular focus on their potentially different impact on health and on ensuring a consistent level of specificity across categories. Broad terms such as recreation were deemed too general to be analytically useful, whereas distinctions like physical versus social activities were considered meaningful. This process ultimately yielded six distinct categories:
physical, mindfulness, nature-appreciation, environmental, social, and cultural activities.

\subsection{Annotating Park OSM Tags with Activities Using LLM Classifiers}
\label{sec:llm-annotations}

To associate different park elements and spaces with health-promoting activities, we annotated \ac{OSM} tags describing those elements and spaces with activities.
This turned out to be a challenging task.
\ac{OSM} is a collaborative platform with some governance and guidelines (\url{https://wiki.openstreetmap.org/wiki/Map_features}) for tagging, but the flexible tagging system offers the crowdsourcing contributors substantial freedom. As a result, the data can be inconsistent and fragmented, necessitating thorough cleaning.
Each map object, such as park elements and spaces, can be tagged with an unlimited number of tags, offering in-depth descriptions. As a result, we encountered over 30,000 unique key–value pairs associated with park elements and spaces.
Since our primary focus is on the core functional aspects of these elements, we conducted a data cleaning step (detailed in Appendix \autoref{app:osm_cleaning}) to remove irrelevant metadata associated with the map objects. This filtering allowed us to focus exclusively on tags relevant to activity-related features, thereby making the annotation process more pertinent to our study.

\subsubsection*{Using Large Language Models for Annotation}

Even for domain experts, linking these tags unequivocally to health-related activities was difficult. For instance, a bench might relate to socializing, enjoying nature, or resting after physical activity. Choosing one activity over another often depended on personal experience, %
as many tags could plausibly refer to multiple activities. Given the large number of items and the specialized nature of the task, we chose an \ac{LLM} classifier as an alternative to expert annotation or crowdsourcing.

Using \acp{LLM} as classifiers offers several advantages as they provide a more objective and consistent approach to annotation, can handle large volumes of data quickly, and do so at relatively low cost. In some cases, \ac{LLM} classifiers have even outperformed crowdworkers\cite{Alizadeh2023Open}, who may themselves rely on machine learning tools to complete tasks\cite{Veselovsky2023Artificial}. Research also suggests that \ac{LLM}-based annotations can match the quality of those made by domain experts\cite{Ziems2023Can}.
Given these benefits, we established a benchmark in our domain to evaluate the feasibility of using \acp{LLM} to map \ac{OSM} tags to health-promoting activities.
Three experts manually annotated the 100 most frequent tags, and final labels were selected by majority vote.
We then used this expert-labeled dataset to assess the accuracy of the labels generated by different LLM classifiers.
The details of the LLM annotation benchmark can be found in the Appendix, \autoref{app:benchmark_llm}.
The outcome indicated that GPT-4, set at a temperature of $0.9$, yielded the best annotation performance of an F$_1$ score of $0.77$.

\subsubsection*{Operationalization of the Taxonomy} 

Using the taxonomy with six categories of health-promoting activities in \autoref{tab:activities_literature_full}, and GPT-4 as the best-performing annotation model, we ran the annotation of \ac{OSM} tags describing \emph{park elements} and \emph{park spaces}. These tags were then labeled with one of the health-promoting activities, or \emph{``none''} if they didn't support a particular activity. 
In doing so, we established a lexicon of \emph{park elements} and \emph{spaces} linked to health-promoting activities (\autoref{tab:activities_osm_tags}). We had to exclude the mindfulness activities category at this stage, as none of the \ac{OSM} tags found in parks were primarily associated with it.

\subsection{Computing Park Health Scores by Aggregating OSM Tags}
\label{sec:scoring_method}
The core method to characterize parks in terms of their potential for health-promoting activities is based on counting the respective \emph{park elements} and \emph{spaces}.
These counts are then combined to give each park an overall score for each health-related category. This score represents the potential health benefits of each park.

\subsubsection*{Counting Health-promoting Elements and Spaces in Parks}

In our process of assigning health-promoting activity scores to each park, we first gathered \emph{park elements} and \emph{spaces} within each park using the \emph{osmium} library. %
We then assigned health-promoting activities to these \emph{elements} and \emph{spaces} based on the lexicon created in the previous step (\autoref{tab:activities_osm_tags}).
We discarded any \emph{elements} or \emph{spaces} whose tags did not match an activity category.
In a few instances, \emph{park elements} or \emph{spaces} could  fall into more than one health-promoting activity category. For example, apple trees are annotated in \ac{OSM} with the tags \texttt{[natural=tree, produce=apple]}. In our lexicon of park elements and spaces, we map \texttt{natural=tree} to the nature-appreciation and \texttt{produce=apple} to the environmental category. To account for this overlap, we proportionally assign the element as 50\% nature-appreciation and 50\% environmental. More generally, when multiple tags are matched to different categories, we proportionally count the resource based on the number of matched tags, ensuring that its contributions are accurately accounted for, thus avoiding underestimating secondary activities of multi-purpose facilities.

\subsubsection*{Transforming Counts into Health Scores}

After tallying up the \emph{park elements} and \emph{spaces} within the park, we measured the overall effect of the park in promoting healthy activities within a city. This score should account for the park's size and the range of facilities it offers for different activities. Our proposed scoring method is based on the following considerations.

\begin{enumerate}
	\item \emph{Amount of Health-Promoting Elements and Spaces:} The number and area of health-promoting facilities determine how much offering there is for each activity.
	\item \emph{Area of the Park:} The character of a park depends on the concentration of health-promoting facilities. Larger parks must offer more to obtain a high score.
	\item \emph{Diminishing Returns with Increased Count:} We posit that as the count of these elements and spaces increases, the associated benefits exhibit diminishing returns.
	\item \emph{City-specific Normalization of Park Health Scores:} The value of a park's facilities for a certain activity is relative to similar facilities in other parks throughout the city.
\end{enumerate}

To reflect these assumptions into our scoring, we proposed a linear regression model to compute the park health scores shown in \autoref{eq:linear_log_models}.
We used the idea of an \emph{``average park''} in each city to compute a baseline and used the distance of each park to the average park line, i.e., the residual, as a score.
The average park baseline was determined by computing separate linear regression models for \emph{park elements} and \emph{spaces} in each city, estimating the expected amount of facilities relative to the park area. 

\begin{equation}
	\label{eq:linear_log_models}
	E_{Act}(log_2(count(Act))) = i + s \cdot log_2(\text{park area}) ~|~Act  \in \text{Activity Categories},
\end{equation}

where $i$ and $s$ represent the intercept and slope of the regression lines, respectively.
To obtain regression models for each activity category and both \emph{park elements} and \emph{park spaces}, we utilized the binary logarithm to account for the diminishing returns of an increase in park size.
For each city, separate regression models were calculated for each activity category, as well as for \emph{park elements} and \emph{park spaces}. See \autoref{app:regression} for details on the modeling.

\begin{figure}[t]
	\includegraphics[width=\textwidth]{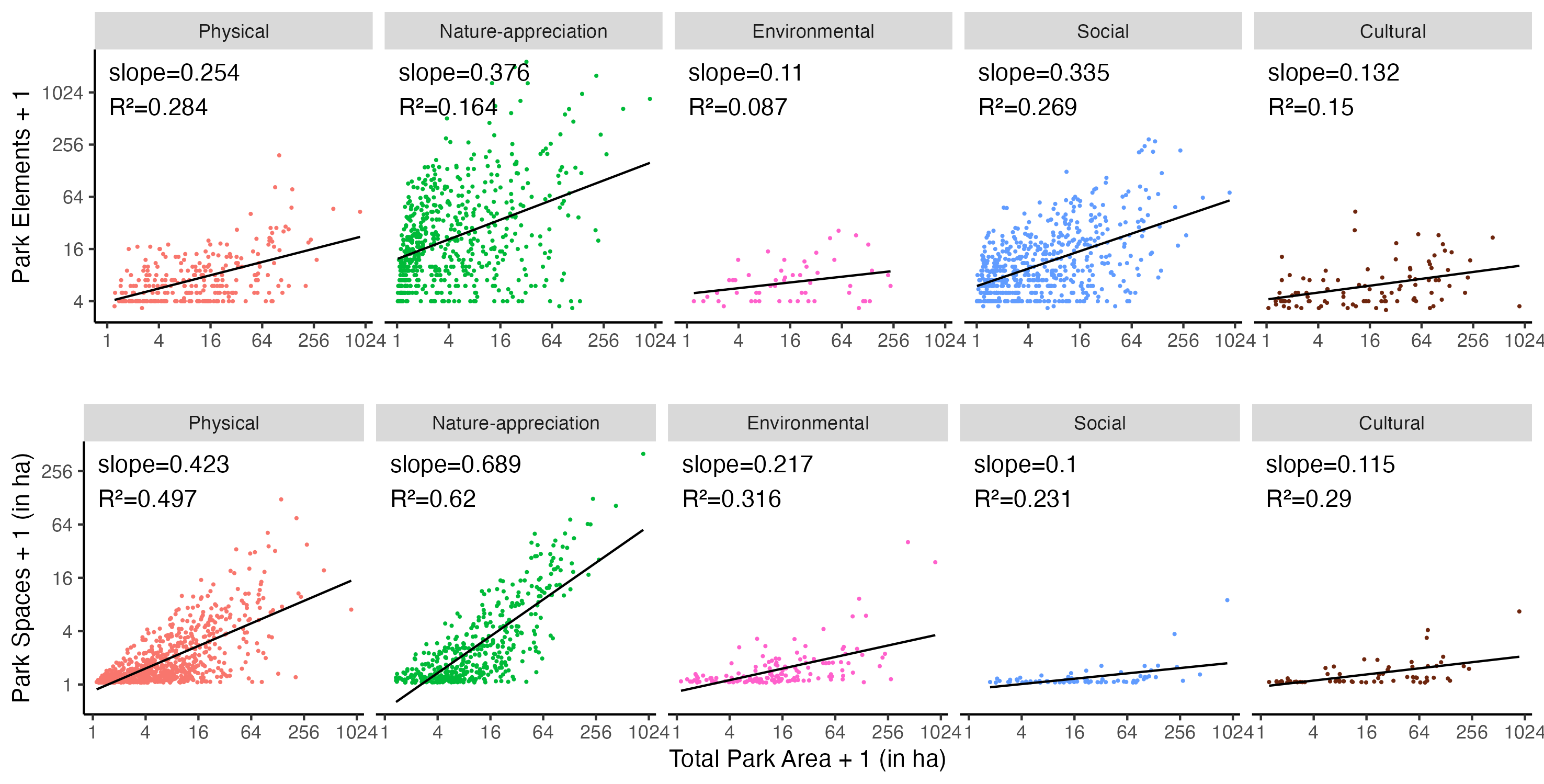}
	\caption{\textbf{Visualization of the scoring method.} The linear regression models for determining the park health scores for the \emph{park elements} (top) and \emph{park spaces} (bottom) in London, UK. The horizontal axis denotes the park's area ($log_2$), while the vertical axis represents the count of \emph{park elements} ($log_2$) and the area occupied by health-promoting \emph{park spaces} ($log_2$). The modest $R^2$ values are anticipated, highlighting the variety among parks.}
	\label{fig:log-log_elements_spaces}
\end{figure}

By analyzing the residuals, we identified parks that exceeded expectations (positive residuals) and those that fell short (negative residuals) in providing health-promoting resources for a given activity.
We made the linear model dependent on park area so that the resulting scores for health-promoting \emph{elements} and \emph{spaces} reflected their density. To reduce the influence of extremely large parks with high amounts of facilities, we applied the binary logarithm.
We calculated these scores separately for each city, rather than using a single global model, to ensure that the results reflected each city's local context.
To illustrate this method, we plotted the linear models and the individual park scores in the $log$--$log$ space for London, UK (\autoref{fig:log-log_elements_spaces}).
The regression lines denote expected health scores based on park size.
The modest $R^2$ values showed that the number of health-promoting facilities could not be explained by park size but instead reflected different design priorities and the needs of local citizens.
Park scores for \emph{park elements} and \emph{spaces} are residuals from this \emph{average park line} in the model, adjusting for park size when determining health scores.

\subsubsection*{Combining Scores from Park Elements and Spaces}

The regression models gave us individual scores for \emph{park elements} and \emph{park spaces} for each activity.
To verify the impact of combining these scores into one combined health score, we examined the co-occurrences of \emph{park elements} and \emph{spaces} and found that they represent orthogonal concepts in practice, as outlined in \autoref{app:orthogonal_elem_spaces}.
Based on this finding, we computed a combined score by first normalizing the scores of \emph{park elements} and \emph{spaces} using the z-score transformation, considering all parks in a city.
This normalization allowed us to standardize the scores, making them comparable despite being on different scales.
Then, we linearly combined these $z-$scores by averaging them together to create the overall score for the park according to \autoref{eq:park_scores}.

\begin{equation}
	\label{eq:park_scores}
	Score(P_{Act}) = \frac{z(residual_{elements}(P_{Act})) + z(residual_{spaces}(P_{Act}))}{2},
\end{equation}
where $P$ denotes an individual park, $Act$ is one of the activity categories,  the residual scores for elements and spaces stem from \autoref{eq:linear_log_models}, and $z()$ indicates the $z-$score transformation.

The combined and normalized scores of \emph{park elements} and \emph{spaces} represent a comprehensive and unified measure of the park's health-promoting amenities and facilities, accounting for both individual elements and cultivated areas.
The combination process accounted for the relative importance of each aspect, leading to a more meaningful overall score that represents how well a park is equipped to support performing health-promoting activities.
Since the scores are based on $z-$score–normalized residuals, a value around $0$ indicates average support for a given activity, while a score of $\pm 1$ means the park is $1$ standard deviation above or below the city-wide average.

\subsection{Quantifying Disparities of Park Scores}
\label{sec:gini_formula}

One goal of our study was to quantify disparities in health benefits offered by different parks within a city.
To measure the disparities in the presence of amenities and facilities associated with health-promoting activities within a city, we propose the following disparity index.
The metric essentially quantifies the inequality of the park health scores, as generally, one could expect that good park management would provide for a similar amount of features and facilities in all parks of a city.
Since the park health scores could be negative, we can not directly use a standard inequality metric, such as the Gini Index, but had to min-max normalize the park score before computing the Gini Index (\autoref{eq:gini_scores}).

\begin{equation}
\label{eq:gini_scores}
Gini_{Act}(\{X_{Act}' | X_{Act} :P \in C\}),
\end{equation}

\noindent where  $X_{Act}$ is the score of activity category ${Act} \in [\text{physical, cultural, etc.}]$ of a park P in city C, and \[X' = \frac{X-X_{max}}{X_{max}-X_{min}} (min-max~normalization),\] and the Gini index was computed in a standard way:
\[Gini = \frac{A}{A + B}, \]
where $A$ is the area between the Lorenz curve and the line of perfect equality and $B$ is the area beyond the Lorenz curve\cite{Barrow2017statistics}.

\subsection{Semantic Matching of Flickr Labels and OSM Tags}
\label{sec:matching_flickr}

In our validation, we leveraged a global dataset of 10.7 million geotagged Flickr images taken within parks across 35 cities. Each image came with user-generated tags, partially annotated by computer vision algorithms. To semantically match these Flickr labels to \ac{OSM} tags, we used text embeddings, treating the task as an asymmetric semantic search problem.
To overcome language diversity in the Flickr labels, we detected the top three non-English languages per city and translated the labels into English using machine translation models.
To further improve embedding quality, we enriched OSM tags with concise definitions from the OSM mapping guidelines. The embeddings were generated using the \texttt{all-mpnet-base-v2} S-BERT model, and matching was done based on cosine similarity, with a threshold of $0.7$ to ensure quality.
The methodological details are described in \autoref{app:flickr_matching}.

This process yielded $2,171$ Flickr-to-OSM matches, with $1,432$ corresponding to health-promoting features.
To assess accuracy, three experts reviewed the $20$ most frequent label-tag matches for London. We aggregated their responses using majority voting. The experts' annotation agreed with $82\%$ of the matchings, which is highly accurate considering they are based solely on individual tags.

Having assured that the matchings are accurate, we proceeded to profile the parks based on the activities associated with the matched \ac{OSM} tags, following the same scoring approach as what we used for the \ac{OSM} \emph{park elements} and \emph{spaces} (\autoref{eq:linear_log_models}).
In our validation, we chose a minimum of $250$ images from each park and at least $15$ parks in each city. This criterion was established to secure a robust number of images for each park, thus ensuring the accuracy of our analysis. This was a mitigation against potential biases that could have been introduced by individual photographers if a park had only a few images.

\subsection*{Data availability}
The replication package contains tables of the park health scores in the cities: \url{https://github.com/LinusDietz/Health-Promoting-Parks-Replication}.
The original OpenStreetMap data used for scoring the parks is publicly available and can be best obtained from one of the third-party download servers, for example from \url{https://download.geofabrik.de}.
The Flickr dataset for the validation can not be shared due to the terms of conditions of this dataset.

%% file: sections/appendix.tex
\appendix

\part*{Appendix}

\section{Supplementary  Results}

\subsection{Lexicon of Elements and Spaces for Health-promoting Activities in Parks}

We display the top 10 most frequent tags for each activity category in \autoref{tab:activities_osm_tags} and have published all 1441 entries of the lexicon in the replication package.

\newcolumntype{x}{>{\raggedright\arraybackslash}X}
\begin{table}[htb]
	\centering
	\small
	\caption{\textbf{Lexicon of \emph{elements} and \emph{spaces} for health-promoting activities in parks.} We show the 10 most frequent tags per activity category. The full lexicon of 1441 elements and spaces is available in the replication package.}
	\label{tab:activities_osm_tags}
	\begin{tabularx}{\textwidth}{>{\raggedright}p{2.45cm}xx}
		\toprule
		\textbf{Activity Category}& \textbf{Elements} & \textbf{Spaces}\\
		\midrule
		\textbf{Cultural} & \input{tables/top_park_elements_cultural.tex} &  \input{tables/top_park_spaces_cultural.tex}\\
		\midrule
		\textbf{Environmental} & \input{tables/top_park_elements_environmental.tex} &  \input{tables/top_park_spaces_environmental.tex}\\
		\midrule
		\textbf{Nature-appreciation} & \input{tables/top_park_elements_nature.tex} &  \input{tables/top_park_spaces_nature.tex}\\
		\midrule
		\textbf{Physical} & \input{tables/top_park_elements_physical.tex} &  \input{tables/top_park_spaces_physical.tex}\\
		\midrule
		\textbf{Social} & \input{tables/top_park_elements_social.tex} &  \input{tables/top_park_spaces_social.tex}\\
		\bottomrule
	\end{tabularx}
\end{table}	

\subsection{Top Parks by City}

The following \autoref{tab:top_city_parks} lists the best-equipped park of each city by their scores on each activity category.

\begin{table}[t!]
	\centering
	\caption{Top parks of the cities by activity category}
	\label{tab:top_city_parks}
	\footnotesize
	\input{tables/top_city_parks.tex}
\end{table}

\subsection{Geographic Influence on Park Scores}
\label{app:geo_influence}

\begin{figure}[htb]
\centering
	\includegraphics[width=.80\textwidth]{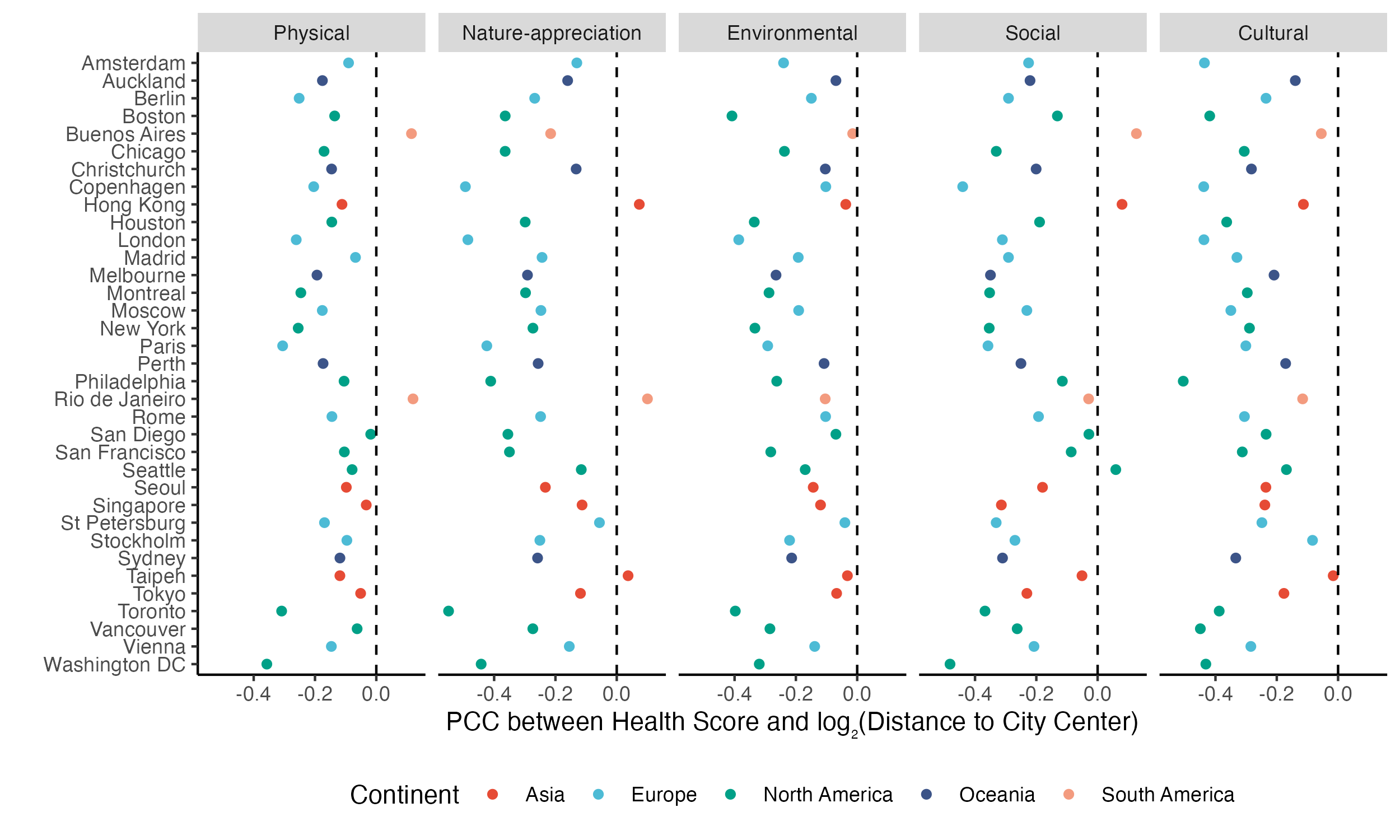}
	\caption{Pearson correlations between park health scores and the distance to the city center ($log_2$) by city.}
	\label{fig:corr_dist_cc}
\end{figure}

\autoref{fig:corr_dist_cc} complements the findings from \autoref{sec:geographic_influence} regarding geographic influences on park health scores.
The plot provides additional insights subdivided by continents.

Using a correlation analysis between the distance to the city center (discounted using the binary logarithm) and the park health scores, we aimed to provide an additional quantification of the decay in park scores moving away from the city center.
Plotting the \ac{PCC} for each city individually in  \autoref{fig:corr_dist_cc}, we observe mostly low to moderate negative correlations.

\begin{table}[!t]
	\centering
	\caption{\textbf{Park health scores by quartiles determined using the distance to the city center}. Q1 are the inner city parks, and Q4 are the parks that are most distant to the city center. On the left, we show the mean health scores for each quartile; on the right, we show the $p$-values determined using a two-sided t-test for each null hypothesis $H_0:  Q_i = Q_j $ stating that the mean scores for two subsequent quartiles $i$ and $j$ are equal. All null hypotheses can be refuted with high significance levels ($p<0.001, \ast\ast\ast$), with the exception of the difference between Q3 and Q4 in the environmental category, where the significance is $p=0.003, \ast\ast$. We confirmed the normal distribution of the scores using QQ-plots.}
	\label{tab:quartiles_categories}
	\input{tables/quartiles_categories.tex}
\end{table}

\subsection{Disparities of Park Scores}

\begin{figure}[htb]
	\includegraphics[width=\textwidth]{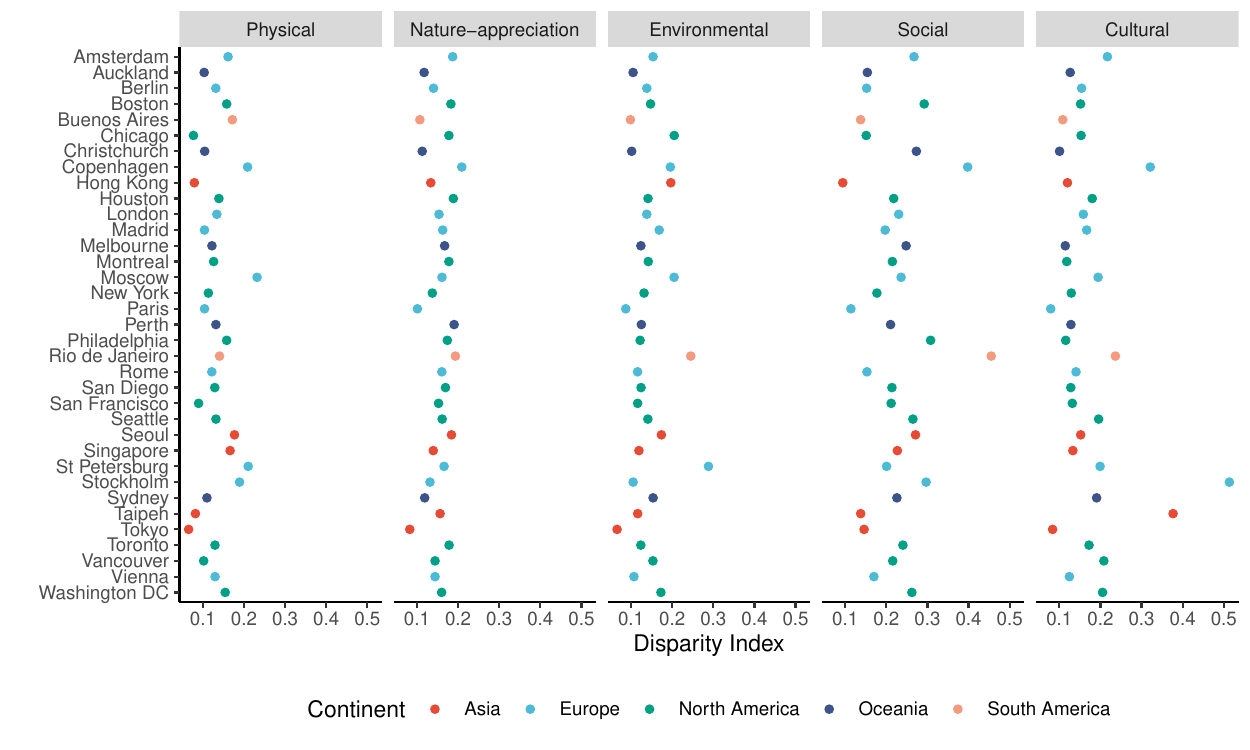}
	\caption{Inequality of park scores in the cities. Overall, nature-appreciation activities show the lowest disparities, whereas the largest differences can be observed in the social activities category.}
	\label{fig:gini_categories}
\end{figure}

\autoref{fig:gini_categories} provides a visual representation of the data tabulated in \autoref{tab:gini_cities} emphasizing the continent of the cities. This figure supplements the findings in \autoref{sec:results_inequality_continents}.

\subsection{Validating the Overall Ranking of Parks through an Online Survey}

As an additional means for validating the park scores, we conducted a survey in one city, London, UK.
In an online questionnaire, we asked London citizens about suitable parks for performing activities.
The main set of questions was phrased as: \textit{``Can you name several parks suitable for \textbf{physical activities} (e.g., sports)?}''

\subsubsection*{Study Information}

We recruited the participants using the first author's institutional research recruitment portal as well as mailing lists within scientific institutions in London.
The participants were informed about the voluntary nature of their participation and that no personally identifiable information about them was collected.
For these reasons, age was collected using 7 age groups (\emph{``Below 18''}, \emph{``18--24''}, \emph{``25--34''}, \emph{``35--44''}, \emph{``45--54''}, \emph{``55--64''}, \emph{``65 and over''}) and as a privacy mechanism only the postal area (e.g., N1) instead of the full postcode was requested.
Furthermore, we asked participants how long they have been living in London (\emph{``I don't live in London.''} -- \emph{``Less than 1 year.''} -- \emph{``1 to 5 years.''} -- \emph{``More than 5 years.''}).
Finally, as a means to identify low-quality responses, we asked people for a park close to their homes, which we could use as an instructional manipulation check in conjunction with the reported postal area.
The data collection was registered as a minimal-risk study at the first author's institutional review board (King's College London Research Ethics Office, ID: \texttt{MRA-22/23-38802}).

\subsubsection*{Results}

The metric we used to quantify how well the citizen response aligned with our health scores is the average percentile-ranking\cite{Quercia2010Recommending,Hu2008Collaborative}, which captures how highly the selected park was placed in the overall ranking of parks for the corresponding activity.
A value close to $1$ means parks with the highest scores were selected, whereas $0.5$ would represent a random selection.
The results demonstrate a clear alignment between the freely recalled parks by the participants and the rankings derived from our health scores.
As shown in \autoref{tab:online_survey_results}, the median and mean values of the average percentile-ranking for the parks named by citizens were consistently high.
For nature-appreciation, physical activities, cultural activities, and social activities the median scores are above $0.89$, highlighting a strong concordance between citizens' perceptions of the park and the quantitative rankings derived from our proposed park profiling method.

\begin{table}[htb]
	\centering
	\caption{\textbf{Result statistics of the online survey.} Citizens were asked to name parks that are suitable for the activities. The first three columns show the statistics of the average percentile-ranking of the named parks. \textbf{AR} is the answer rate of the respective category, i.e., how many respondents were able to name at least one park), \textbf{N} is the number of non-empty responses, and \textbf{MR} is the mean number of parks that were named per respondent.}
	\label{tab:online_survey_results}
	\input{tables/online_survey_overall.tex}
\end{table}

The result for \ environmental activities is subpar compared to the other activities, with a mean and median average percentile-ranking of $0.50$.
Only 57.5\% of the respondents could name an \ environmental park, and on average, $1.8$ parks were named in this category by each person, which indicates that parks for environmental activities are harder to think of compared to the other activities.
Another explanation for the low scores in this category is that while urban gardening and conservation can be done in many parks, they typically do not occupy much \emph{spaces},  environmental activities are less mainstream in cities, and the number of \emph{park elements} for this category is comparatively low in London's parks impeding high activities scores.

The overall alignment between Londoners' perceptions of parks and our health scores underscores the effectiveness of our approach in accurately capturing and evaluating the health-promoting potential of parks. 
\begin{figure}[htb]
	\includegraphics[width=\textwidth]{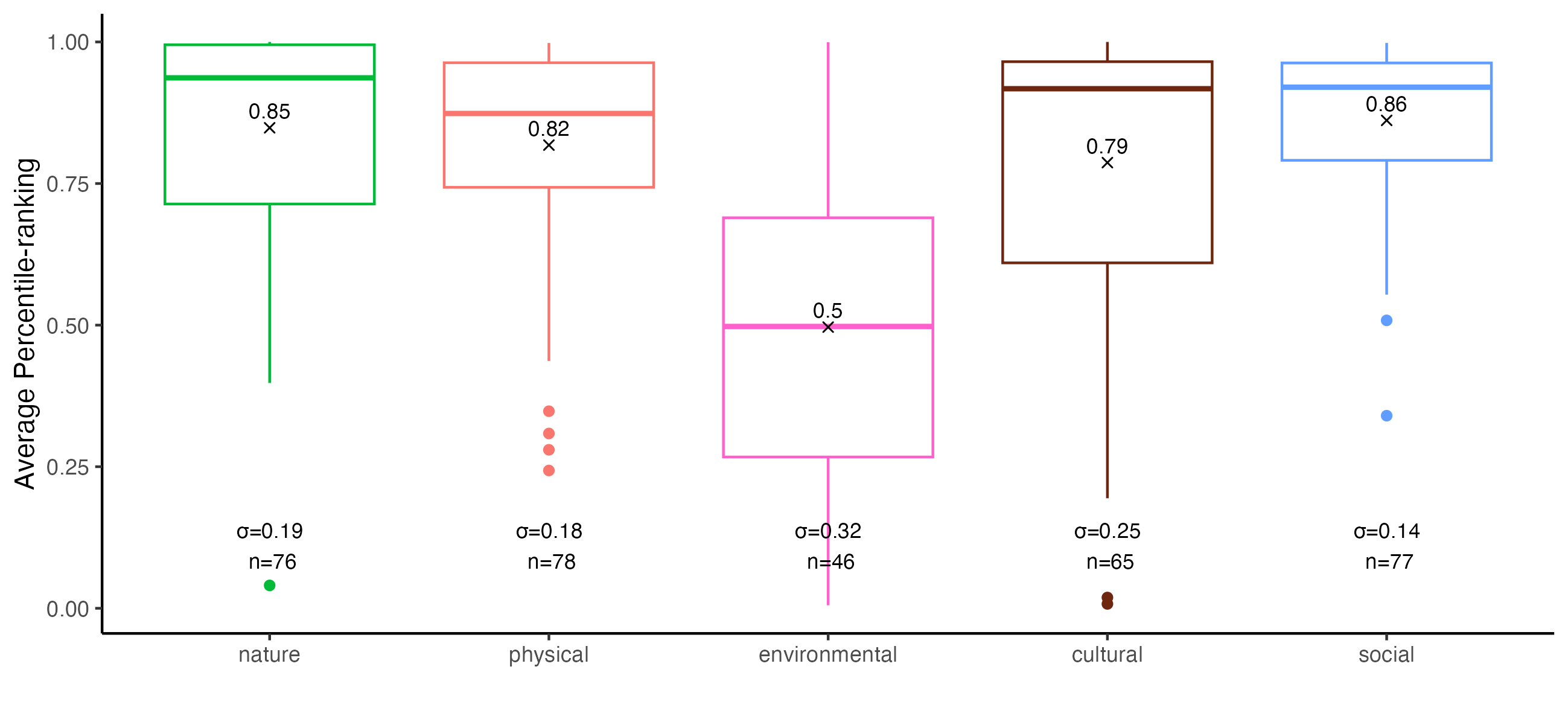}
	\caption{The online survey results as box-whisker plots. \boxplotdef The values indicated with the x represent the mean values.}
	\label{fig:online_survey_results_distribution}
\end{figure}

\section{Supplementary Material}

\subsection{Scoping Review to Map Activity Categories to Health Benefits}

Next, our goal was to collate and map health-prompting activities in parks discussed in prior studies.
Considering between a systematic and scoping type of review, the scoping review was a better fit for our task because we only needed to map activities discussed in the literature, and we did not need to focus on the types and quality of data collected in those studies, which is a task for systematic reviews.
Specifically, we turned to using the well-established PRISMA method\cite{Page2021PRISMA}, which is designed to facilitate transparent reporting of reviews, and it has been designed primarily for reviews of studies that evaluate the effects of health interventions, irrespective of the design and strength of effects found in the included studies. 

The overarching research question was: \emph{``Which are the health benefits of activities in urban greenery?''}
Our focus on urban greenery instead of only parks was to ensure both the \emph{comprehensiveness} and \emph{generality} of the taxonomy, as future studies might look beyond urban parks.
We used the WHO's definition of urban greenery to determine the scope of our survey: \emph{``[\ldots] urban green space is defined as all urban land covered by vegetation of any kind. This covers vegetation on private and public grounds, irrespective of size and function, and can also include small water bodies such as ponds, lakes, or streams (``blue spaces'')\cite{WHO2017Urban}.}''

As we were interested in the intersection of urban greenery and medical studies, we performed a set of queries on PubMed and SpringerLink to identify papers that linked the usage of urban greenery with health benefits. An article was deemed relevant if the results evidenced that one or more activities typically done in public urban green spaces had a health benefit.
To obtain a comprehensive overview of each activity category, we used a total of 6 queries.
Upon our preliminary experiments, we employed a collection of keywords for our queries that included both those commonly encountered in the initial set of studies and those formulated by our experts. This approach enabled us to discover a diverse range of papers relevant to each category of activity. The queries were: 

\begin{description}
	\item \textbf{Physical activities:} (urban greenery) AND (health) AND (sports OR exercise)
	\item \textbf{Nature-appreciation activities:}  (urban greenery) AND (health) AND (nature) AND (exposure)
	\item \textbf{Environmental activities:}  (urban greenery) AND (health) AND (garden OR planting OR conservation)
	\item \textbf{Social activities:}  (urban greenery) AND (health) AND (social OR social cohesion OR social capital OR social contacts)
	\item \textbf{Cultural activities:}  (urban greenery) AND (health) AND (culture) OR (cultural ecosystem)
		\item \textbf{Mind-body activities:}  (urban greenery) AND (health) AND (mindfulness OR meditation OR yoga OR tai chi OR breathing techniques)
\end{description}

\begin{figure}[htb]
	\centering
	\includegraphics[width=1\columnwidth]{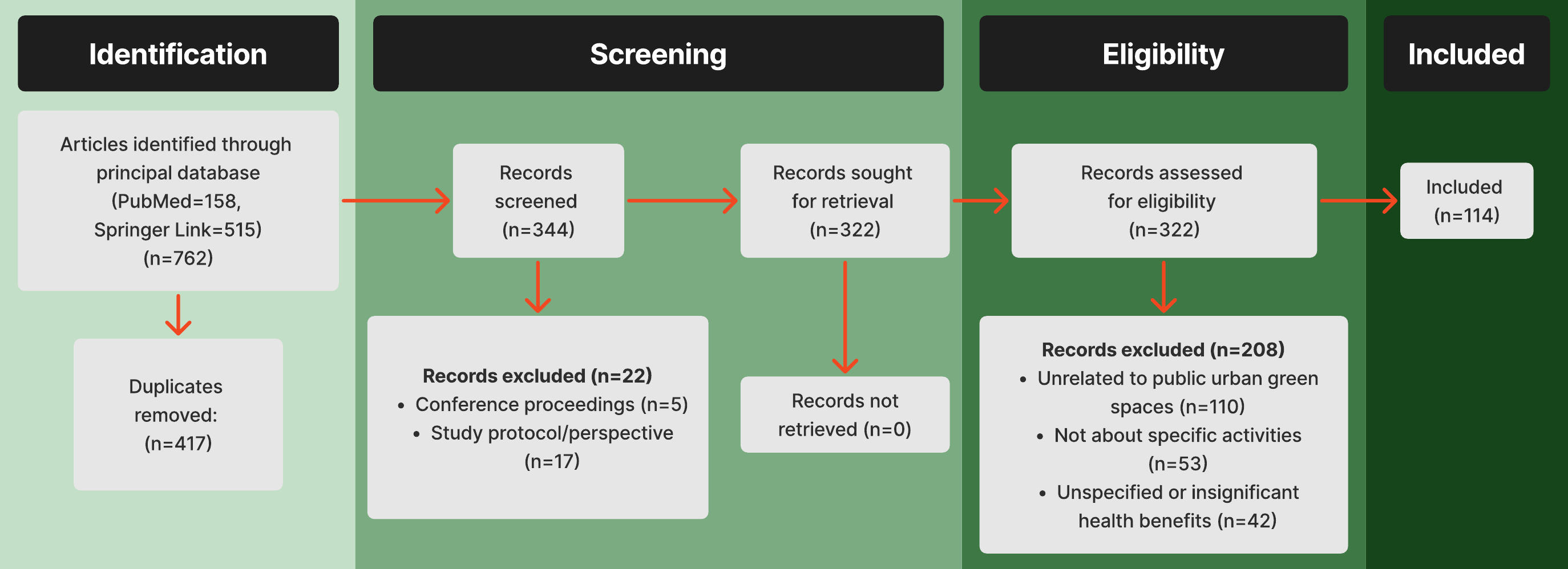}
	\caption{\textbf{Our PRISMA Statement:} Process of identification, screening, and determining eligibility for articles in our literature survey.}
	\label{fig:prisma_flow}
\end{figure}

Following the PRISMA statement depicted in \autoref{fig:prisma_flow}, out of the initially \emph{identified} 762 articles, 417 were duplicates, leaving us with 344 unique articles.
Next, we \emph{screened} these articles and discarded 5 conference proceedings and 17 articles that were perspectives or study protocols, successfully retrieving the remaining 322 articles. In the \emph{Eligibility} step, we determined whether these articles were relevant to our search. We found that 114 articles were relevant, while 208 were not. Most articles were excluded because they were not about urban green spaces or because there was no significant link between the activities and health benefits.
When analyzing the \textit{included} articles, we recorded each activity category alongside the general health aspects and specific health benefits the article evidenced (\autoref{tab:activities_literature_full}).

\subsubsection*{Results}

\begin{table}[!t]
	\footnotesize
	\centering
	\caption{\textbf{Activities in urban parks linked to health benefits.} Specific health benefits evidenced in the respective articles are grouped by health aspects. }
	\label{tab:activities_literature_full}
	\input{tables/literature_activities_benefits_all.tex}
\end{table} 

A can  be seen in \autoref{tab:activities_literature_full}, most research has focused on the benefits of \emph{physical activity} in parks. Out of 79 studies on the health benefits of exercising, 46 underscored positive outcomes like weight loss\cite{10.3390/ijerph15102186,10.3389/fpubh.2023.1207975,10.1123/jpah.2012-0503}, cardiovascular improvements\cite{10.1016/j.envres.2021.112449,10.1007/BF03391647}, metabolic activity\cite{10.1186/s12966-017-0625-5,10.3390/ijerph16162948}.
Additionally, these activities demonstrated positive effects on mental health (16 articles), well-being (7 articles), and social health (6 articles).

The second most studied category is \emph{nature-appreciation}, with 68 articles. These activities significantly boost mental health (34 articles), primarily in reducing stress\cite{10.1186/1471-2458-6-149,10.1016/j.envint.2021.106664} and anxiety\cite{10.1038/s41370-021-00349-x,10.1177/19375867211059757} and preventing depression\cite{10.1038/srep28551,10.1186/1471-2458-12-337,10.3390/ijerph14020172,10.1177/19375867211059757,10.1016/j.bpsgos.2022.01.004,10.3390/ijerph18105137,10.3389/fpsyt.2022.757056}. They also contribute to physical health (14 articles) and overall well-being (12 articles).

In our review, we found that \emph{social} and \emph{environmental activities} received less attention in conjunction with urban parks, with only 33 and 28 articles covering them, respectively. Despite this, both contribute to all identified health aspects. Social activities enhance social and mental health, fostering a sense of belonging\cite{10.1007/s10393-014-0939-6,10.3389/fpubh.2022.1105473,10.1016/j.envres.2018.09.033} and improving mood\cite{10.1140/epjds/s13688-021-00278-7}. Environmental activities, such as gardening, offer diverse benefits, including cognitive restoration\cite{10.3390/ijerph15081705} and improved general health\cite{10.1186/s12889-020-08762-x}.

Finally, \emph{cultural} and \emph{mindfulness activities} are relatively under-researched regarding their health benefits in the context of urban parks.
Cultural activities often fell outside the scope of our review, which required a connection to urban parks, while more general cultural activities were studied. However, their health benefits are likely underreported given the presence of cultural facilities in parks (e.g., historic monuments or arts venues).
Likewise, mindfulness activities and nature-based therapeutic interventions, such as forest therapy\cite{ROSA2021126943} provide health benefits, but have not been studied in the context of urban parks, highlighting a potential gap in the scientific literature that warrants future exploration.

In summary, we found that all the different activities we identified have a \emph{distinct but overlapping set of health benefits}. For example, both physical and nature-appreciation activities help prevent depression, but only physical activities help bone development, and only nature-appreciation brings calmness.

\subsection{Determining the Threshold Values for Computing the Linear Models}
\label{app:thresholds}

\begin{figure}[htb]
	\includegraphics[width=\textwidth]{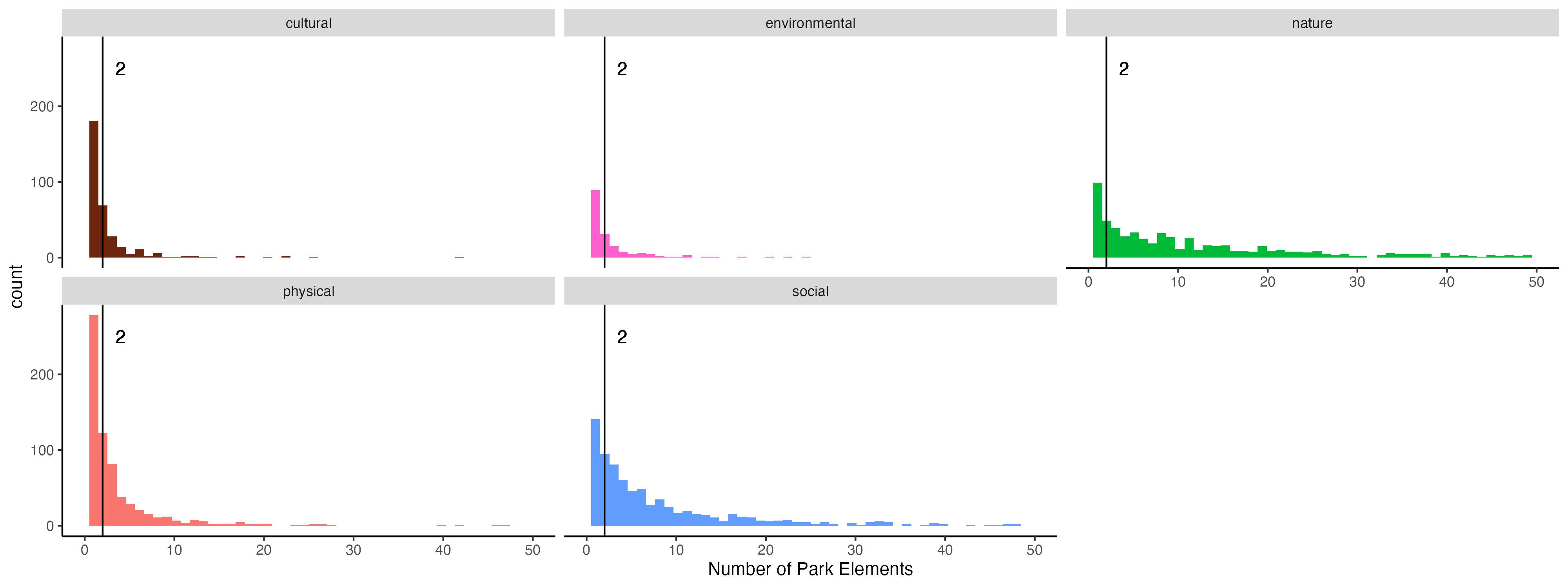}
	\caption{\textbf{Histogram of park elements.} We set 2 as the minimum number per activity category.}
	\label{fig:th_elements}
\end{figure}

\begin{figure}[htb]
	\includegraphics[width=\textwidth]{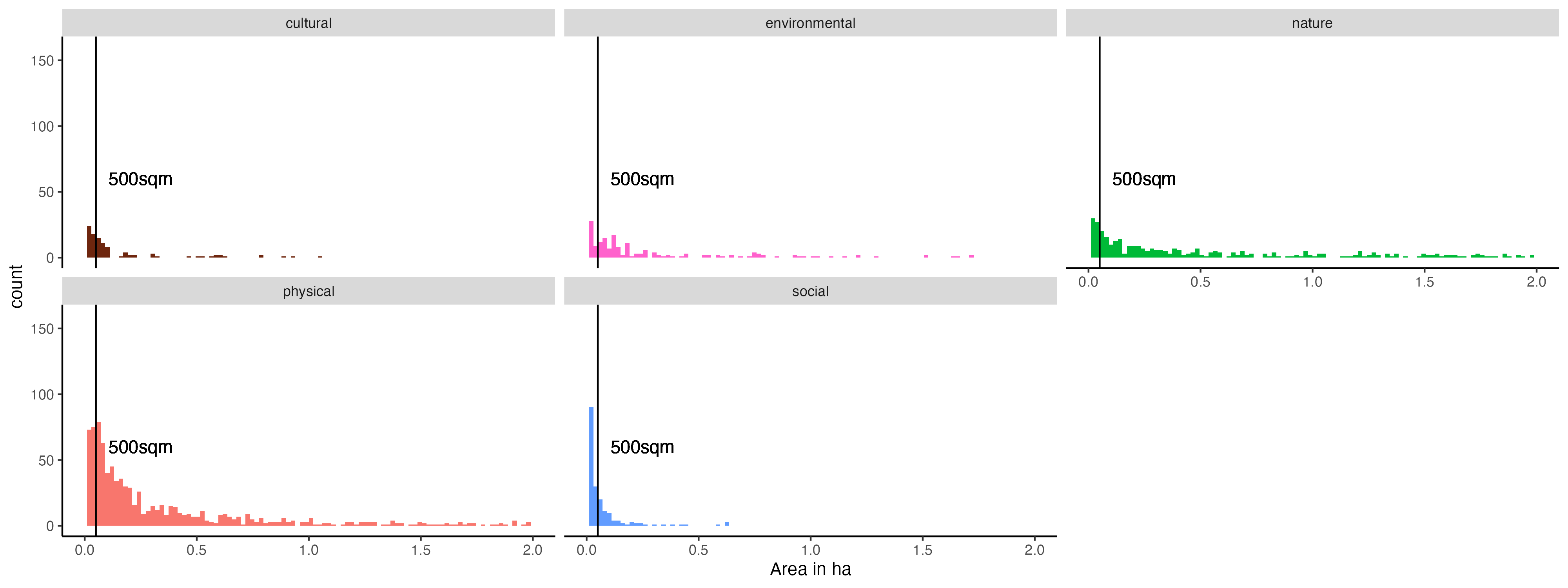}
	\caption{\textbf{Histogram of park spaces.} We set 0.05ha (500 m$^2$) as the minimum size per activity category.}
	\label{fig:th_spaces}
\end{figure}

\autoref{fig:th_elements} and \autoref{fig:th_spaces} depict histograms  of park elements and park spaces. The plots supplement the determination of thresholds for excluding parks with insufficient activity data in \autoref{sec:scoring_method}, \emph{Transforming Counts to Health Scores}.

\subsection{Data Cleaning of OSM Tags}
\label{app:osm_cleaning}

This section describes the steps we undertook to exclude OSM tags that are not useful for our analysis. 

The first step of the data cleaning process was not specific to annotating health-promoting activities. Instead, the focus was on removing any extra information not necessary for understanding the main purpose of the map object. For example, to identify a bench on a map, one just needs to look for the label \texttt{amenity=bench}. However, a bench can also have additional tags like \texttt{inscription}, \texttt{operator}, \texttt{material}, and \texttt{backrest}, which offer more specifics about the bench. When it comes to identifying the object's primary purpose for health-promoting activities, this extra information is not only unnecessary but could also lead to confusion. 
To remove these irrelevant labels, three of the authors created lists of keys and values that were only used to provide extra details when combined with other labels. All co-authors carefully reviewed, discussed, and agreed upon these lists. If there was any doubt about whether to exclude certain labels, they were kept and left for subsequent annotation.
The goal was to make sure that only necessary and relevant labels were kept for categorizing \emph{park elements} and \emph{spaces} into health-promoting activity categories.

In the process of cleaning the data, 1926 keys were omitted.
These included keys such as \texttt{name}, \texttt{operator}, and \texttt{source}, which cannot provide insight into the object's activity.
In addition, 11 values were also left out because they only described metadata and did not help in understanding the primary function or essence of the map object.
Examples of such values include \texttt{yes/no}, \texttt{unknown}, or \texttt{Bing}.
A full list of these omitted keys and values can be found in the replication repository.
This initial data cleaning step significantly reduced the number of tags to 2118, which were the ones we needed to map to health-promoting activities, or none if the object did not support any of them.
This streamlined dataset provided a more focused and relevant basis for the subsequent annotation and classification of \emph{park elements} and \emph{spaces}.

\subsection{Benchmarking LLM Classifiers}
\label{app:benchmark_llm}

To evaluate the suitability of \ac{LLM} classifiers as annotators for \ac{OSM} tags, we created a high-quality, expert-annotated gold standard set consisting of the 100 most frequent tags.
To ensure accuracy and reliability, three co-authors independently labeled these 100 items with health-promoting activities or none, and we used the majority voting strategy to aggregate the individual opinions into one final outcome label.
In cases where conflicts arose, i.e., where the three annotators provided different labels, a discussion was held to resolve the discrepancies.
Through this rigorous annotation process, we established a robust and reliable \emph{``gold standard''} dataset of 100 items.
This dataset serves as a benchmark to assess the accuracy of the labels provided by the \ac{LLM} classifiers.

For generating the annotations, we conducted a systematic exploration of the configuration settings of two \acp{LLM}, GPT-3.5-turbo\cite{brown2020language} and GPT-4\cite{openai2023gpt4} using the OpenAI API\cite{OpenAIAPI}.
Our goal was to identify the best-performing setting in terms of the weighted $F_1$ score, which is the harmonic mean of precision and recall in this multi-class classification task.
The independent variables were 
	\emph{i)} the \acl{LLM}, i.e., \texttt{gpt-3.5-turbo} or \texttt{gpt-4}\cite{openai2023gpt4},
	\emph{ii)} the temperature parameter $t \in \{0.3; 0.6; 0.9\}$, which controls the randomness of the models' completions, and
	\emph{iii)} the prompt, for which we tested two versions, one with and without providing a brief definition of the \ac{OSM} tag taken from the \ac{OSM} wiki. The full prompt is shown in \autoref{fig:llm_prompt}.

\begin{figure}[htb]
	\fbox{%
		\parbox{\textwidth}{
			{\color{gray}$\Rightarrow$ You are an expert in urban planning and public health, with a specialization in urban parks. You have studied how parks promote health and have an understanding of the various activities that people engage in within them. Proficient in the OpenStreetMap project and skilled in tagging urban elements, particularly those within parks, your responsibility involves assigning activities to specific park elements based on OpenStreetMap tags.}
			
			$\Rightarrow$ Consider these 6 categories of activities people do in urban parks:
			\begin{description}
				\item[Physical activities] This category is about leisure pursuits that involve physical movement and sports. Example activities are:  Walking, hiking, trail running, biking, swimming, rock climbing, canoeing, kayaking, horseback riding, outdoor sports, and group fitness classes.
				\item[Mind-body activities] This category is about physical practices that combine movement, breathing techniques, and meditation to promote relaxation, stress reduction, and overall well-being. Example activities are: Yoga, meditation, and tai chi.
				\item[Nature appreciation activities] This category is about leisure pursuits that involve enjoying and exploring the natural world. Example activities are: Bird watching, camping, picnicking, fishing, painting, drawing, photography, and nature journaling.
				\item[Environmental activities] This category is about gardening and conservation of parks. Example activities are: Gardening, planting trees and flowers, and participating in conservation efforts and volunteering.
				\item[Social activities] This category is about coming together and communal experiences that involve engaging with others. Example activities are: Attending outdoor festivals and concerts, playing sports, and volunteering.
				\item[Cultural activities] This category is about activities that celebrate and showcase the cultural diversity and heritage of the community. Example activities may include cultural festivals, art exhibits, music performances, storytelling, and cultural workshops. 
			\end{description}
			Given the categories of "Physical activities", "Mind-body activities", "Nature appreciation activities", "Environmental activities", "Social activities", "Cultural activities", and "none of the above"; How would you classify the following OpenStreetMap Tags found in a park and which [0,100] reliability scores (even multiple scores) would you give to your assessment? You can assign up to two categories, but you don't have to!
			
			{\color{gray}$\Rightarrow$ Provide the answer in the format:
				Main Activity$\mid$Reliability Score (of the main activity)$\mid$Secondary Activity (if it exists)$\mid$Reliability Score (of the secondary activity)$\mid$Explanation (not more than 20 words)}
			
			$\Rightarrow$ leisure=playground\\
			$\Rightarrow$ \ul{Social$\mid$80$\mid$Physical$\mid$70$\mid$Children utilize playgrounds for both physical activity and exercise, as well as for social interactions through cooperative play.}
			
			$\Rightarrow$ leisure=pitch\\
			$\Rightarrow$ \ul{Physical$\mid$90$\mid$Social$\mid$75$\mid$People engage in team sports on sports pitches, encompassing both physical activity and a social aspect due to the collaborative nature of team-based sports.}
			
			$\Rightarrow$ leisure=garden\\
			$\Rightarrow$ \ul{Nature-appreciation$\mid$80$\mid$Environmental$\mid$60$\mid$People utilize gardens to cultivate an appreciation for nature, while gardens also serve as spaces for environmental activities, such as planting flowers and engaging in gardening practices.}
			
			$\Rightarrow$ amenity=bench\\
			$\Rightarrow$ \ul{Social$\mid$80$\mid$Nature-appreciation$\mid$40$\mid$People utilize benches for relaxation and social interaction, fostering a sense of community. Moreover, benches offer the opportunity to immerse oneself in the surrounding nature, encouraging moments of contemplation and introspection.}
			
			$\Rightarrow$ natural=wood\\
			$\Rightarrow$ \ul{Nature-appreciation$\mid$100$\mid$None$\mid$100$\mid$Forested areas, such as woods, primarily serve the purpose of appreciating and connecting with nature.}
			
			$\Rightarrow$ amenity=parking\_space\\
			$\Rightarrow$ \ul{None$\mid$90$\mid$None$\mid$100$\mid$A parking space does not inherently cater to a specific activity.}
	}}
	\caption{Preparatory prompt provided to the \ac{LLM} classifiers via the OpenAI API. 
		The tag and the definition were subsequently prompted. Regular text refers to `user' messages, gray text refers to `system' messages, and underlined text refers to `assistant' messages. $\Rightarrow$ denotes the beginning of a new message. Bold markup was added for improved readability.}
	\label{fig:llm_prompt}
\end{figure}

\autoref{fig:llm_prompt} shows a specific sequence of prompts designed to elicit a main activity and a secondary activity for each \ac{OSM} tag.
The reason behind this approach was our hypothesis that certain \ac{OSM} tags could support multiple health-promoting activities, as demonstrated by the example of benches that could be argued to be annotated with social, nature-appreciation, or physical activities.
Additionally, we obtained a reliability score for each of the model's annotations.
These reliability scores offer an indication of the model's confidence in its assigned activities, which could serve as a threshold to actually use the annotations, as low scores might indicate that the annotation is more speculative.
By incorporating these main activities, secondary activities, and reliability scores from the \ac{LLM} models, we hoped to gain a more nuanced insight into how these amenities and facilities in parks can be used.
This detailed information allowed us to account for the potential multi-functionality of certain \ac{OSM} tags and provided data for the evaluation using the proposed benchmark.

Furthermore, we followed the guidelines\cite{openai2024Guidelines} to optimize the performance of the \acp{LLM} annotations.
We assigned a system persona, i.e., \emph{`You are an expert in urban planning and public health, with a specialization in urban parks. [\ldots]'}, gave definitions of the six activities with exemplary activities, and provided several correct completions of items as means to few-shot learning.
Finally, we provided a clear specification of the desired output format.

To determine the highest agreement between the human-annotated benchmark and the annotations of the \acp{LLM}, we used the $F_1$ score, which is the harmonic mean of the precision and recall.
One complication in the evaluation was that the benchmark only comprised one activity label, whereas we asked the LLM annotator for a main and secondary activity for each tag.
Thus, we report two $F_1$ scores: one that uses the label from the main category only and another that is a weighted combination of the main activity category and the secondary activity category.
The weighted $F_1'$ score is computed by slightly altering the impact of each element of the confusion matrix as follows:

\begin{equation}
	TP' = {TP^i_{main} \cdot reliability^i + TP^i_{2nd} \cdot (1-reliability^i)} \hfill \text{(true positives)}
\end{equation}

\begin{equation}
	FP' = {FP^i_{main} \cdot reliability^i + FP^i_{2nd} \cdot (1-reliability^i)}  \hfill \text{(false positives)}
\end{equation}

\begin{equation}
	FN' = {FN^i_{main} \cdot reliability^i + FN^i_{2nd} \cdot (1-reliability^i)}  \hfill \text{(false negatives)}
\end{equation}

\begin{equation}
	reliability^i = \frac{mean(reliability^i_{main})}{mean(reliability^i_{main}) +mean(reliability^i_{2nd})}   \hfill \text{(Ratio of reliability between main and secondary category)}
\end{equation}

Intuitively, this means that we use the reliability scores stemming from the LLM annotations to estimate the LLM's confidence that a label is correct, thus creating a comparable metric that allows for comparing two annotations for one item to one human-annotated ground truth.

\begin{table}[t!]
	\centering
	\caption{Results of \ac{LLM} Benchmarking. The highest performance is achieved with GPT-4, using a temperature of 0.9 and not providing definitions for the tags. Annotating a secondary activity did not improve the $F_1$ scores.}
	\input{tables/LLM_benchmark.tex}
	\label{tab:llm-benchmark}
\end{table}

We tested various settings to see which would deliver the best performance, which was GPT-4, set at a temperature of 0.9, and without providing definitions for tags. 
To give you a clearer picture, we've compiled the results of the top-performing configuration in \autoref{tab:llm-benchmark}.
The F$_1$ scores tabulated in the tables show the best results of systematically adjusting the reliability scores for primary and secondary categories from 0 (using any label, regardless of its reliability) to 1 (annotate ``none'' in all cases). 
Generally, GPT-4 outperformed its predecessor, GPT-3.5. Adding definitions actually had a negative effect on label quality, possibly due to misleading keywords in the tagging instructions. When it came to the temperature setting, there was no consistent impact, with minimal differences between otherwise equivalent configurations.
Interestingly, adding a secondary activity label didn't improve the annotation quality (cf. last column of \autoref{tab:llm-benchmark}). In fact, the best results were achieved when the reliability threshold of the secondary annotation was close to 1, rendering all secondary annotations to \emph{``none''}, thus being equivalent to only using the main activity label.
This suggests that the primary labels generated by the system are already of high quality, so putting any weight on a secondary label actually harms the overall score. 
Based on these findings, we decided to use GPT-4, set at a temperature of 0.9 and without definitions, to label all OSM tags and not impose any threshold on the reliability score.

\subsection{Modeling Average Park Offerings}
\label{app:regression}

To model average park offerings, we used independent regression models for \emph{park elements} and \emph{spaces}.
During the computation of the regression models, parks with very low activity counts in a specific category were excluded. 
This exclusion was necessary to prevent artificially flattening the regression lines due to close-to-zero values, which would distort the normalization.
The specific threshold for excluding parks with low activity counts was determined empirically by analyzing the histograms of the values.
This approach enabled us to identify an appropriate cutoff point for excluding parks with insufficient activity data, ensuring the reliability of the regression models.
For a visual representation of the exclusion process and the determination of the threshold, refer to \autoref{fig:th_elements} and \autoref{fig:th_spaces}.

\subsection{Orthogonality of Park Elements and Spaces}
\label{app:orthogonal_elem_spaces}
\emph{Park elements} include points of interest, individual trees, benches, and similar items. \emph{Park spaces}, however, include areas like forests, sports fields, and buildings.
There can be cases where a park area is broken down into its individual parts, like a playground with separately mapped features like swings, slides, or spinning equipment. But these cases are pretty rare in \ac{OSM} mapping. Likewise, unless a tree is particularly important, areas tagged as \texttt{natural=wood}  should not include individual trees according to the mapping guidelines\cite{OpenStreetMapWiki2024Map}.
Based on these observations, we hypothesized that it would be acceptable to combine scores from park features and areas linearly, as they contribute differently to the overall offering of facilities for health-promoting activities.

To validate the assumption that \emph{park spaces} and \emph{park elements} are orthogonal, we calculated the pairwise correlation coefficients of their respective scores in all cities and averaged them, as presented in \autoref{tab:corr_elements_spaces}.
The low Pearson Correlation Coefficients supported our observation that the scores of \emph{park spaces} and \emph{park elements} indeed capture largely independent concepts, with all correlations being slightly positive but below $0.2$.
Consequently, we combined them into one overall score for the park.

\begin{table}[t!]
	\centering
	\caption{\textbf{The correlations between the  scores for \emph{park elements} and \emph{park spaces} are low.} This property allowed us to linearly combine them into a unified score.}
	\label{tab:corr_elements_spaces}
	\input{tables/no_residual_correlations.tex}
\end{table}

\subsection{Semantic Matching of Flickr Labels and OSM Tags}
\label{app:flickr_matching}
In our validation, we used a global dataset of geotagged photos from Flickr, from which we selected all 10,711,513 images that were taken within one of the parks from 35 cities in our study.
These images came with user-generated labels and were also partially annotated with computer vision labels from a computer vision algorithm\cite{Li2009Towards,Thomee2016YFCC100M}.
To obtain semantically equivalent representations of Flickr labels and \ac{OSM} tags, we employed Sentence-BERT (S-BERT\cite{Reimers2019Sentence}) for text embeddings.
We formulated this task as an asymmetric semantic search problem, where the Flickr label was the search term, and the goal was to find the closest matching OSM tag.
Given the worldwide reach of our study, the multiple languages present in the user-generated Flickr labels created a challenge in mapping them to the corresponding OSM tags, which were all in English.
To address this, we identified the top three languages besides English used in the tags of each city, using the Google MediaPipe\cite{Lugaresi2019MediaPipe} Language Detection Model\cite{MediaPipeAuthors2023Language}.
To ensure that the language detection was accurate and to eliminate named entities, we only used labels where the language detection indicated a confidence of 50\% or more.
We then translated those tags to English using the respective OPUS machine translation models\cite{Tiedemann2020OPUS}.

To further improve the quality of embeddings, we augmented the \ac{OSM} tags with short definitions sourced from the \ac{OSM} mapping guidelines\cite{OpenStreetMapWiki2024Map}.
For instance, the  \ac{OSM} tag \texttt{sport=table\_tennis} was augmented with the definition \emph{"A bat and ball game played over a table."}
We were able to expand 66\% of the OSM tags with these descriptions. The remaining tags were left without descriptions primarily because of the unregulated nature of tagging in OSM, which led to many undocumented tags or multiple values within one tag, like \texttt{sports=soccer;rugby}. Note that these tags were still used for mapping, albeit with less information.

After embedding the \ac{OSM} tags using S-BERT's \texttt{all-mpnet-base-v2} model, we proceeded to match each Flickr label to the closest \ac{OSM} tag in the embedding space, using the cosine distance as similarity measure.
To ensure that the matches were of high quality, we set a strict threshold: the cosine similarity score had to be at least 0.7. We arrived at this value after noticing that when the similarity score was lower than 0.7, the matches became less reliable based on manual inspections. 
This allowed us to avoid matching labels that did not have meaningful OSM counterparts. For example, abstract labels describing certain phenomena like \emph{``cloud'',} \emph{``rain'',} and \emph{``sunset''} were not matched.

A detailed review of the matched pairs revealed that, as anticipated, most pairings were logical based on the text similarity between labels and tags with definitions. However, some minor adjustments were still needed, as some matches were not entirely consistent with the theme of health-promoting activities in parks. For example, the term \emph{``outdoor''} was initially linked to \texttt{swimming\_pool=outdoor}. But as there cannot be a suitable equivalent for \emph{``outdoor''} on OSM, we removed this pairing and equivalent ones, such as \emph{``park,''} as all photos were taken in parks. 
Another instance was the pairing of ``water'', which did not capture the specific role of water features in parks in promoting health. We manually adjusted this to \texttt{water=river}, which better reflects bodies of water commonly found in parks.
Through this review step, we improved the quality of the matched pairs, ensuring they more closely align with the theme of health-promoting park activities. The need for this manual step should not diminish the effectiveness of the semantic search within sentence embeddings. It was merely to eliminate labels that could not meaningfully correspond with an OSM tag and to match a few labels with more domain-relevant tags. 
This matching process yielded 2,171 label-tag pairs in total. Of these, 1,432 pairs corresponded to an OSM tag with health-promoting benefits, such as \emph{``steeplechase''} being matched to \texttt{athletics=steeplechase} involving physical health benefits, while 739 pairs, such as \emph{``Lamp Post''} being matched to \texttt{man\_made=lamp\_post} did not imply health benefits.

We evaluated the accuracy of the resulting label-tag matchings by asking three domain experts to independently assess whether the 20 most frequent matchings from Flickr tags to activity categories were plausible and correct. We aggregated their responses using majority voting. Given the multiple languages present in the dataset, we used only the tags from London in this evaluation step, as they were in English. The experts agreed with 82\% of the matchings, which is highly accurate considering they are based solely on individual tags.

Having assured that the matchings are accurate, we proceeded to profile the parks based on the activities associated with the matched \ac{OSM} tags, following the same scoring approach as what we used for the \ac{OSM} \emph{park elements} and \emph{spaces} (cf. \autoref{eq:park_scores}).
In our validation, we chose a minimum of 250 images from each park and at least 15 parks in each city. This criterion was established to secure a robust number of images for each park, enhancing the accuracy of our analysis. This approach helped us avoid any potential bias that could have been introduced by individual photographers if a park had only a few images.

\begin{figure}[htb]
	\centering
	\includegraphics[width=\textwidth]{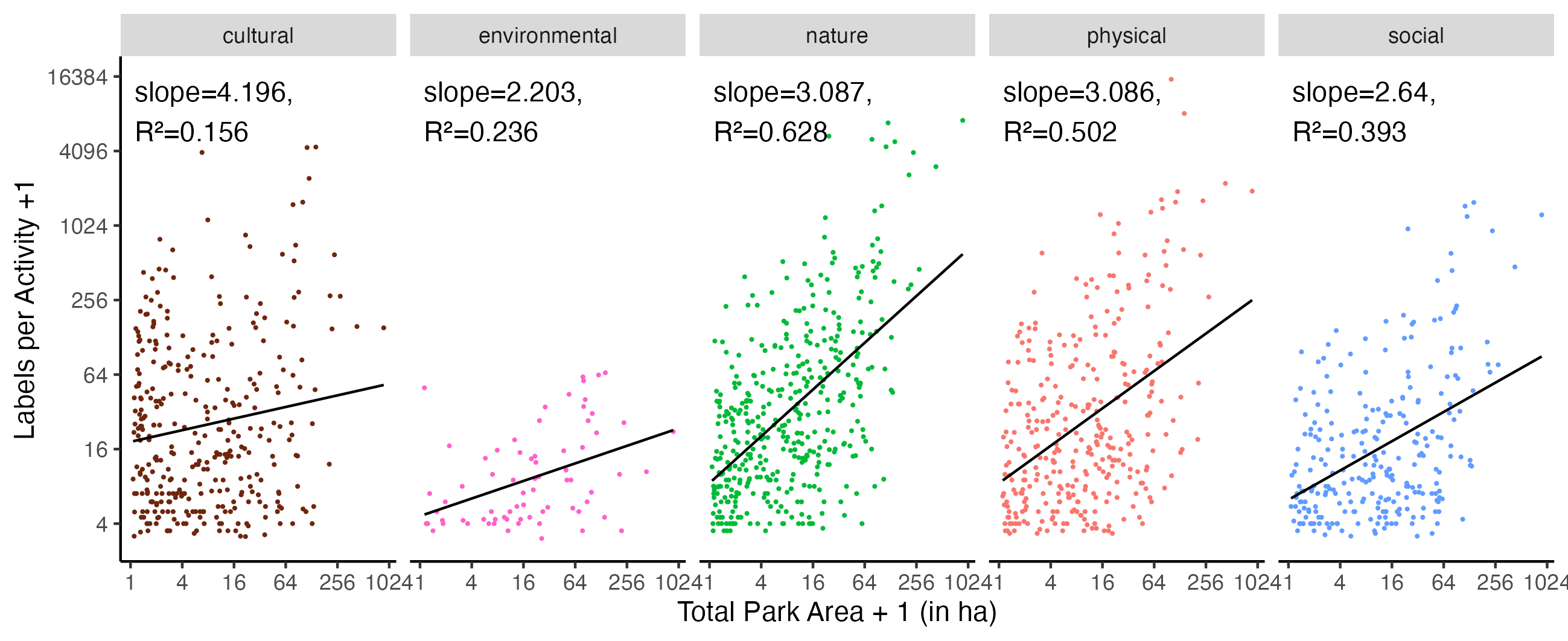}
	\caption{Visualizing the fit of the linear model for determining the park scores using matched Flickr labels for London, UK. The horizontal axis denotes the park's area ($log_2$), and the vertical axis is the number of categorized labels of images from these parks ($log_2$).}
	\label{fig:log_log_flickr}
\end{figure}

\autoref{fig:log_log_flickr} depicts the computation of the Flickr activity scores. The method is the same as for the \ac{OSM} tags; however, on the y-axis, we use the count of the matched Flickr labels instead of the \ac{OSM} tags.
The method for scoring is described in \autoref{sec:scoring_method}; the mapping of Flickr Labels to OSM tags and activities is explained in \autoref{sec:matching_flickr}.

\subsection{User Studies with Urban Designers and Park Maintenance Experts}
\label{sec:expert_rubric}

We conducted three semi-structured interviews, each lasting about 30 minutes, with urban designers and park maintenance experts. Participants were recruited through our network and by direct email to municipalities. The experts' backgrounds were as follows:

\begin{itemize}
    \item Lecturer and urban designer (E1). 20 years of practical and academic experience in Hong Kong and London.
    \item  Urban designer and master planner (E2). 30 years of experience in the United States, United Kingdom, Europe, and the Middle East.
    \item  Parks development manager at a local council (E3). 12 years of experience as conversationalist, arboricultural manager in diverse contexts.
\end{itemize}

The online interviews consisted of 4 steps.

\begin{enumerate}
    \item \textbf{Introduction}. Three questions to understand their experience in urban park design and management.
    \item \textbf{Study brief.} A summary of our study and main findings.
    \item \textbf{Visualization.} Interacting with the visualization and exploring a chosen city.
    \item \textbf{Debriefing.} Open-ended questions and feedback on the project.
\end{enumerate}

We qualitatively analyzed the interview transcripts~\cite{miles1994qualitative,braun2006using} (recorded with consent) to understand the implications of our work for professional practice, verify inter-cultural validity, and to further validate the taxonomy of activities.

%% file: tables/top_park_elements_cultural.tex
information=board,
 tourism=artwork,
 historic=memorial,
 artwork\_type=sculpture,
 artwork\_type=statue,
 board\_type=history,
 historic=monument,
 memorial=plaque,
 memorial=war\_memorial,
 memorial=bench

%% file: tables/top_park_spaces_cultural.tex
tourism=artwork,
 tourism=attraction,
 religion=christian,
 leisure=bandstand,
 historic=memorial,
 denomination=anglican,
 building=church,
 historic=building,
 tourism=museum,
 amenity=theatre

%% file: tables/top_park_elements_environmental.tex
waste=trash,
 produce=plum,
 amenity=recycling,
 fruit=apple,
 produce=damson,
 leisure=garden,
 man\_made=beehive,
 produce=apple,
 amenity=watering\_place,
 man\_made=monitoring\_station

%% file: tables/top_park_spaces_environmental.tex
leisure=garden,
 landuse=flowerbed,
 landuse=allotments,
 building=greenhouse,
 landuse=orchard,
 landuse=farmland,
 building=farm\_auxiliary,
 landuse=farmyard,
 garden:type=community,
 garden:type=residential

%% file: tables/top_park_elements_nature.tex
natural=tree,
 amenity=fountain,
 tourism=viewpoint,
 board\_type=nature,
 tourism=picnic\_site,
 amenity=shelter,
 natural=shrub,
 attraction=animal,
 board\_type=wildlife,
 waterway=weir

%% file: tables/top_park_spaces_nature.tex
natural=wood,
 natural=water,
 natural=scrub,
 water=pond,
 natural=heath,
 landuse=forest,
 heath=bracken,
 natural=grassland,
 natural=wetland,
 amenity=shelter

%% file: tables/top_park_elements_physical.tex
amenity=bicycle\_parking,
 highway=crossing,
 amenity=drinking\_water,
 leisure=fitness\_station,
 barrier=cycle\_barrier,
 sport=fitness,
 leisure=pitch,
 sport=orienteering,
 orienteering=marker,
 leisure=playground

%% file: tables/top_park_spaces_physical.tex
leisure=pitch,
 leisure=playground,
 sport=soccer,
 sport=tennis,
 highway=footway,
 golf=bunker,
 sport=basketball,
 highway=pedestrian,
 area:highway=footway,
 golf=tee

%% file: tables/top_park_elements_social.tex
amenity=bench,
 tourism=information,
 leisure=picnic\_table,
 amenity=cafe,
 board\_type=notice,
 amenity=telephone,
 amenity=fast\_food,
 playground=playhouse,
 amenity=restaurant,
 advertising=board

%% file: tables/top_park_spaces_social.tex
amenity=cafe,
 building=pavilion,
 building=retail,
 amenity=community\_centre,
 leisure=outdoor\_seating,
 amenity=school,
 amenity=restaurant,
 building=kiosk,
 amenity=kindergarten,
 building=terrace

%% file: tables/top_city_parks.tex
\begin{tabularx}{\textwidth}{lXXXXX}
  \toprule
\textbf{City} & \textbf{Physical} & \textbf{Nature-appreciation} & \textbf{Environmental} & \textbf{Social} & \textbf{Cultural} \\ 
  \midrule
Amsterdam & Sloterpark & Sloterpark & Park Frankendael & Westerpark & Amstelpark \\ 
  Auckland & Lloyd Elsmore Park & Point England Park & Paneke / Radonich Park & Ambury Regional Park & Albert Park \\ 
  Berlin & Tempelhofer Feld & Landschaftspark Johannisthal/Adlershof & Gärten der Welt & Tempelhofer Feld & Treptower Park \\ 
  Boston & Franklin Park & Charles River Esplanade & Temple Street Park & Charles River Esplanade & Georges Island \\ 
  Buenos Aires & Parque Indoamericano & Parque de la Memoria & Paseo Arzoumanian & Parque de las Ciencias & Parque Avellaneda \\ 
  Chicago & Lincoln Park & Northerly Island & Grant Park & Lincoln Park & Grant Park \\ 
  Christchurch & Canterbury Agricutural Park & Bottle Lake Forest Park & Hagley Park North & Avon River Precinct & Hagley Park North \\ 
  Copenhagen & Fælledparken & Østre Anlæg & Husum Bypark & Enghaveparken & Østre Anlæg \\ 
  Hong Kong &\begin{CJK}{UTF8}{min} 九龍仔公園 \end{CJK}Kowloon Tsai Park &\begin{CJK}{UTF8}{min} 藝術公園\end{CJK}Art Park &\begin{CJK}{UTF8}{min} 佐敦谷公園 \end{CJK}Jordan Valley Park &\begin{CJK}{UTF8}{min} 迪欣湖活動中心 \end{CJK}Inspiration Lake Recreation Centre & \begin{CJK}{UTF8}{min}灣仔臨時海濱花園 \end{CJK}Wan Chai Temporary Promenade \\ 
  Houston & Hermann Park & Hermann Park & Wright-Bembry Park & Hermann Park & Hermann Park \\ 
  London & Old Deer Park & Russia Dock Woodland & Bushy Park & Richmond Park & Alexandra Park \\ 
  Madrid & Parque Agustín Rodríguez Sahagún & Dehesa de la Villa & Parque del Retiro & Finca Vista Alegre & Parque del Retiro \\ 
  Melbourne & Albert Park & Grant Reserve & Fitzroy Gardens & Albert Park & Carlton Gardens \\ 
  Montreal & Parc Jean-Drapeau & Parc Angrignon & Jardin botanique de Montréal & Vieux-Port & Parc Jean-Drapeau \\ 
  Moscow & \foreignlanguage{russian}{Парк Останкино} & \foreignlanguage{russian}{Бирюлёвский дендропарк} & \foreignlanguage{russian}{Выставка достижений народного хозяйства} & \foreignlanguage{russian}{Тимирязевский парк} & \foreignlanguage{russian}{Выставка достижений народного хозяйства} \\ 
  New York & Pelham Bay Park & Prospect Park & Bronx Park & Brooklyn Bridge Park & Fort Tilden \\ 
  Paris & Bois de Boulogne & Bois de Vincennes & Jardin des Plantes & Bois de Boulogne & Bois de Boulogne \\ 
  Perth & Altone Park & Kings Park & Hyde Park & Christ Church Grammar Playing Fields & Victoria Gardens \\ 
  Philadelphia & East Fairmount Park & East Fairmount Park & East Fairmount Park & Race Street Pier & Fort Mifflin \\ 
  Rio De Janeiro & Aterro do Flamengo & Quinta da Boa Vista & Largo da Carioca & Campo de Santana & Praça Luís de Camões \\ 
  Rome & Villa Borghese & Villa Glori & Parco Agricolo di Casal del Marmo & Riserva Naturale dell'Acquafredda & Villa Borghese \\ 
  San Diego & Balboa Park & Mission Bay Park & Balboa Park & Balboa Park & Balboa Park \\ 
  San Francisco & Presidio of San Francisco & Lake Merced Park & Golden Gate Park & Presidio of San Francisco & Golden Gate Park \\ 
  Seattle & Warren G. Magnuson Park & Seward Park & Washington Park Arboretum & Seattle Center & Seattle Center \\ 
  Seoul & \begin{CJK}{UTF8}{mj}올림픽공원\end{CJK} & \begin{CJK}{UTF8}{mj}매봉산공원\end{CJK} & \begin{CJK}{UTF8}{mj}서울숲\end{CJK} & \begin{CJK}{UTF8}{mj}송파나루공원\end{CJK} & \begin{CJK}{UTF8}{mj}경복궁\end{CJK} \\ 
  Singapore & Changi Business Park & Windsor Nature Park & Singapore Botanic Gardens & The Lawn@Marina Bay & Singapore Botanic Gardens \\ 
  St Petersburg & \foreignlanguage{russian}{парк Героев-Пожарных} & \foreignlanguage{russian}{Парк-дендрарий Ботанического сада Петра Великого} & \foreignlanguage{russian}{Летний сад} & \foreignlanguage{russian}{Приморский парк Победы} & \foreignlanguage{russian}{парк Лесотехнической академии} \\ 
  Stockholm & Årstafältet & Kungsträdgården & Sveaplan & Karlaplan & Humlegården \\ 
  Sydney & Centennial Park & Centennial Park & Sydney Park & Lawrence Hargrave Reserve & Clarkes Point Reserve \\ 
  Taipeh &\begin{CJK}{UTF8}{min} 大安森林公園\end{CJK} &\begin{CJK}{UTF8}{min} 關渡自然公園\end{CJK} &\begin{CJK}{UTF8}{min} 士林官邸公園\end{CJK} &\begin{CJK}{UTF8}{min} 天母運動公園\end{CJK} & \begin{CJK}{UTF8}{min}中正紀念公園\end{CJK} \\ 
  Tokyo &\begin{CJK}{UTF8}{min} 若洲海浜公園\end{CJK} & \begin{CJK}{UTF8}{min}上野恩賜公園\end{CJK} &\begin{CJK}{UTF8}{min}北の丸公園\end{CJK} & \begin{CJK}{UTF8}{min}上野恩賜公園\end{CJK} &\begin{CJK}{UTF8}{min} 上野恩賜公園\end{CJK} \\ 
  Toronto & Centennial Park & Sunnybrook Park & Highland Creek Ravine & Toronto Island Park & Don Valley Brick Works Park \\ 
  Vancouver & Connaught Park & Stanley Park & Stanley Park & Hastings Park & Morton Park \\ 
  Vienna & Augarten & Draschepark & Schlosspark Schönbrunn & Schlosspark Schönbrunn & Schlosspark Schönbrunn \\ 
  Washington DC & East Potomac Park & National Mall & National Mall & National Mall & National Mall \\ 
   \bottomrule
\end{tabularx}

%% file: tables/quartiles_categories.tex
\begin{tabular}{lrrrrrrr}
  \toprule & \multicolumn{4}{c}{Mean Health Scores} &  \multicolumn{3}{c}{$p$-values of $H_0$} \\ \cmidrule(lr){2-5} \cmidrule(lr){6-8}
Activity Category & Q1 & Q2 & Q3 & Q4 & Q1 = Q2 & Q2 = Q3 & Q3 = Q4 \\ 
  \midrule
    Physical & 0.382 & 0.297 & 0.160 & 0.068 & 0.000 & 0.000 & 0.000 \\ 
  Nature-appreciation & 0.107 & -0.211 & -0.497 & -0.580 & 0.000 & 0.000 & 0.000 \\ 
  Environmental & 0.020 & -0.076 & -0.156 & -0.186 & 0.000 & 0.000 & 0.003 \\ 
  Social & 0.289 & 0.137 & -0.004 & -0.088 & 0.000 & 0.000 & 0.000 \\ 
  Cultural & 0.260 & 0.096 & 0.010 & -0.022 & 0.000 & 0.000 & 0.001 \\ 
   \bottomrule
\end{tabular}

%% file: tables/online_survey_overall.tex
\begin{tabular}{lrrrlrr}
  \toprule
{\textbf{Activity category}} & {\textbf{median}} & {\textbf{mean}} & {\textbf{$\sigma$}} & {\textbf{AR}} & {\textbf{N}} & {\textbf{MR}} \\ 
  \midrule
 Physical & 0.91 & 0.84 & 0.17 & 97.5\% &  78 & 4.26 \\ 
 Nature-appreciation & 0.95 & 0.85 & 0.19 & 95\% &  76 & 2.75 \\ 
 Environmental & 0.50 & 0.50 & 0.31 & 57.5\% &  46 & 1.80 \\ 
 Social & 0.93 & 0.87 & 0.13 & 96.2\% &  77 & 4.94 \\ 
 Cultural & 0.89 & 0.75 & 0.25 & 81.2\% &  65 & 2.27 \\ 
   \bottomrule
\end{tabular}

%% file: tables/literature_activities_benefits_all.tex
\begin{tabularx}{\textwidth}{p{2.1cm}lX}
	\toprule
	\textbf{Activity Category} & \textbf{Health Aspect }&  \textbf{Specific Health Benefit} \\
\midrule
\textbf{Physical} & Cognitive health &                                                                                                                                                                                                                                                                                                                                                                                                                                                                                                                                                                                                                                                                                                                                                                                                                                                                                                                                                                                                                                                                                                                                                                                                                                                                                                                                                                                                                                                              dementia prevention\cite{10.1002/gps.5626} \\
& General health &                                                                                                                                                                                                                                                                                                                                                                                                                                                                                                                                                                                                                                                                                                                                                                                                                                                                                                                                                                                                                                                                                                                                                                                                                                                                                                                                                                                                                                        longevity\cite{10.1007/s10389-018-0956-y,10.1136/jech.56.12.913} \\
& Mental health &                                                                                                                                                                                                                                                                                                                                                                                                                                                                                                                                                                                                                                                                                                                                                                                                                                                                                                                                         stress reduction\cite{10.1016/j.healthplace.2020.102381,10.1016/j.envint.2021.106664,10.1007/s11524-019-00407-8,10.1016/j.socscimed.2013.06.030,10.3390/ijerph182211746}, depression prevention\cite{10.1186/s12889-019-7171-9,10.1186/1471-2458-12-337,10.3390/ijerph14020172,10.1016/j.envres.2022.114081,10.3390/ijerph18105137,10.1038/srep28551,10.3390/ijerph182211746,10.1080/15622975.2022.2112074}, anxiety reduction\cite{10.1016/j.envres.2022.114081}, various\cite{10.1186/s12966-015-0188-2}, mood improvement\cite{10.1136/bjsports-2012-091877} \\
& Physical health &  weight reduction\cite{10.3390/ijerph15102186,10.3389/fpubh.2023.1207975,10.1123/jpah.2012-0503,10.1186/s12966-015-0228-y,10.1007/s10389-018-0956-y,10.1186/s12966-021-01203-x,10.1186/1471-2458-14-233,10.1016/j.ypmed.2021.106790}, increase of physical activity\cite{10.3390/ijerph192416403,10.1186/s12966-017-0625-5,10.3390/ijerph16162948,10.3390/ijerph17176130,10.1016/j.healthplace.2022.102790,10.1186/1479-5868-11-31,10.1016/j.socscimed.2018.05.022,10.3390/ijerph19148848,10.1016/j.socscimed.2013.06.030,10.3390/ijerph18105137,10.1371/journal.pone.0218247,10.1186/s12942-020-00216-2,10.3390/ijerph15061154,10.1016/j.envint.2021.106664,10.1186/1479-5868-11-79,10.1186/s12966-022-01399-6,10.1186/s12889-021-10259-0,10.3390/ijerph17176127,10.1016/j.ufug.2021.127136,10.1186/s12877-022-03679-z,10.1186/s12889-021-10177-1,10.3390/ijerph16081340,10.3389/fpubh.2022.1012222,10.1186/1479-5868-8-125,10.3390/ijerph14020172,10.1186/s12966-021-01088-w,10.1186/s12889-019-7416-7}, blood pressure reduction\cite{10.1038/srep28551}, diabetes prevention\cite{10.1007/s10389-018-0956-y,10.1016/j.envres.2021.112449}, various\cite{10.1186/s12966-015-0188-2}, increase of leisure activities\cite{10.3390/ijerph19159237}, hypertension\cite{10.1007/s10389-018-0956-y}, cardiovascular health improvements\cite{10.1016/j.envres.2021.112449,10.1007/BF03391647,10.1186/s12966-021-01203-x,10.1016/j.puhe.2013.01.004,10.1186/s12966-020-00941-8}, bone development\cite{10.1186/s12966-021-01203-x} \\
& Social health &                                                                                                                                                                                                                                                                                                                                                                                                                                                                                                                                                                                                                                                                                                                                                                                                                                                                                                                                                                                                                                                                                                                                                                                                                                                                                                                 various\cite{10.1016/j.ufug.2020.126888,10.1186/s12966-015-0188-2}, social cohesion\cite{10.1002/ajcp.12559,10.3390/ijerph14020172,10.1016/j.healthplace.2016.08.011,10.1038/srep28551} \\
& Well-being &                                                                                                                                                                                                                                                                                                                                                                                                                                                                                                                                                                                                                                                                                                                                                                                                                                                                                                                                                                                                                                                                                                                                                                                                                                    increase restorative capacity\cite{10.1007/s11524-017-0191-9}, enhanced social interactions\cite{10.3390/ijerph19031466}, quality of life\cite{10.1186/s12889-021-10177-1,10.3390/ijerph15061154,10.3390/ijerph18105137,10.1002/gps.5626,10.1186/s12966-020-00941-8} \\
\midrule

\multirow{2}{2cm}{\textbf{Nature-appreciation}}& Cognitive health &                                                                                                                                                                                                                                                                                                                                                                                                                                                                                                                                                                                                                                                                                                                                                                                                                                                                                                                                                                                                                                                                                                                                                                                                                                                                                                                                                                                                                                            attention fatigue reduction\cite{10.3389/fpubh.2023.1218091} \\
       & General health &                                                                                                                                                                                                                                                                                                                                                                                                                                                                                                                                                                                                                                                                                                                                                                                                                                                                                                                                                                                                                                                                                                                                                                                                                                                                                                                                                                                                                                                        lower morbidity\cite{10.1186/s12889-020-08762-x} \\
       & Mental health &                                                                                                                                                                                                                                                                                                                                                             positive emotions\cite{10.3390/ijerph17020394,10.1038/s41598-022-24637-0}, depression prevention\cite{10.1038/srep28551,10.1080/13607863.2018.1516193,10.1186/1471-2458-12-337,10.3390/ijerph14020172,10.1177/19375867211059757,10.1016/j.bpsgos.2022.01.004,10.3390/ijerph18105137,10.3389/fpsyt.2022.757056}, anxiety reduction\cite{10.1038/s41370-021-00349-x,10.1177/19375867211059757,10.1016/j.bpsgos.2022.01.004,10.1186/s12889-021-11622-x}, suicide prevention\cite{10.1016/j.envint.2020.105982,10.1016/S2542-5196(18)30033-0}, mood improvement\cite{10.1007/s11524-023-00757-4,10.1136/bjsports-2012-091877,10.3389/fpsyt.2022.757056,10.3389/fpubh.2023.1218091}, relaxation\cite{10.1038/s41598-022-24637-0}, mindfulness\cite{10.1038/s41598-022-24637-0}, calmness\cite{10.1186/s12889-021-11622-x}, stress reduction\cite{10.1186/1471-2458-6-149,10.1016/j.envint.2021.106664,10.3390/ijerph191711059,10.1186/s12889-020-8300-1,10.3389/fpubh.2023.1218091,10.3390/ijerph19169778,10.1177/19375867211059757,10.1016/j.envres.2021.111233,10.1016/j.bpsgos.2022.01.004,10.1007/s11524-023-00757-4,10.3390/ijerph16224393} \\
       & Physical health &                                                                                                                                                                                                                                                                                                                                                                                                                                                                                                                                                                                                                                                                                                                                                                                                                                                                                   mood improvement\cite{10.1007/s11524-023-00757-4}, improved ghq-12 scores\cite{10.1016/j.envres.2018.06.004}, blood pressure reduction\cite{10.1007/s11524-023-00757-4,10.1186/1471-2458-7-118,10.1038/srep28551}, antenatal health\cite{10.1007/s11906-020-01088-4}, respiratory health\cite{10.1016/j.scitotenv.2022.154447}, increase of physical activity\cite{10.3389/fpubh.2022.1012222,10.1186/s12889-023-16243-0,10.1016/j.envint.2021.106664,10.1016/j.envres.2018.06.004,10.3390/ijerph14020172,10.1007/s40279-018-0917-1,10.3390/ijerph18105137}, blood oxygen saturation\cite{10.1007/s11524-023-00757-4} \\
       & Social health &                                                                                                                                                                                                                                                                                                                                                                                                                                                                                                                                                                                                                                                                                                                                                                                                                                                                                                                                                                                                                                                                                                                                                                                                                                                                increased social capital\cite{10.1016/j.healthplace.2017.11.006}, social loneliness reduction\cite{10.1007/s00127-022-02381-0}, various\cite{10.1016/j.ufug.2020.126888}, social cohesion\cite{10.3390/ijerph14020172,10.1038/srep28551} \\
       & Well-being &                                                                                                                                                                                                                                                                                                                                                                                                                                                                                                                                                                                                                                                                                                                                                                                                                                                                                                                                                                                                                                                       stress reduction\cite{10.1177/19375867211059757,10.1016/j.envres.2021.111233,10.1007/s11524-023-00757-4}, quality of life\cite{10.1080/13607863.2018.1516193,10.3390/ijerph15102180,10.1016/j.envres.2019.108535,10.1177/19375867211059757,10.1016/j.envres.2021.111233,10.3390/ijerph18105137,10.1186/s12877-016-0288-0}, blood pressure reduction\cite{10.1007/s11524-023-00757-4}, increase restorative capacity\cite{10.3390/ijerph182312628} \\
              \midrule

\textbf{Environmental} & Cognitive health &                                                                                                                                                                                                                                                                                                                                                                                                                                                                                                                                                                                                                                                                                                                                                                                                                                                                                                                                                                                                                                                                                                                                                                                                                                                                                                                                                                                                                              restorative effect against cognitive failures\cite{10.3390/ijerph15081705} \\
& General health &                                                                                                                                                                                                                                                                                                                                                                                                                                                                                                                                                                                                                                                                                                                                                                                                                                                                                                                                                                                                                                                                                                                                                                                                                                                                                                                                                                                                                                                        lower morbidity\cite{10.1186/s12889-020-08762-x} \\
& Mental health &                                                                                                                                                                                                                                                                                                                                                                                                                                                                                                                                                                                                                                                                                                                                                                                                                                                                                                                                                                                                                                                                                                                                                                                                  stress reduction\cite{10.3390/ijerph19169778,10.1016/j.scitotenv.2017.11.160}, anxiety reduction\cite{10.1038/s41598-020-74828-w,10.3390/rs12081350}, improved sleep\cite{10.1038/s41598-020-74828-w,10.3390/rs12081350}, depression prevention\cite{10.3390/rs12081350,10.1080/13607863.2018.1516193} \\
& Physical health &                                                                                                                                                                                                                                                                                                                                                                                                                                                                                                                                                                                                                                                                                                                                                                                                                                                                                                                                                                                                                                                                     cardiovascular health improvements\cite{10.1016/j.envint.2019.105181}, inflammation reduction\cite{10.3390/ijerph15081705}, respiratory health\cite{10.1016/j.envint.2019.105181}, access to healthy produce\cite{10.1111/inm.13149}, immune system improvement\cite{10.1093/gerona/glaa271}, increase of physical activity\cite{10.3390/ijerph16162948,10.1186/s12889-018-5812-z}, improved sleep\cite{10.1038/s41598-020-74828-w} \\
& Social health &                                                                                                                                                                                                                                                                                                                                                                                                                                                                                                                                                                                                                                                                                                                                                                                                                                                                                                                                                                                                                                                                                                                                                                                                                                                                                                                                                                              access to healthy produce\cite{10.1111/inm.13149}, social cohesion\cite{10.1016/j.scitotenv.2017.11.160,10.1111/inm.13149} \\
& Well-being &                                                                                                                                                                                                                                                                                                                                                                                                                                                                                                                                                                                                                                                                                                                                                                                                                                                                                                                                                                                                                                                                                                                                                                                                 nutritional diversity\cite{10.1186/s13002-020-00421-0}, quality of life\cite{10.1038/s41598-020-74828-w,10.1186/s12889-015-2574-8,10.1186/s12889-018-5661-9,10.1080/13607863.2018.1516193}, improved sleep\cite{10.1038/s41598-020-74828-w}, increase restorative capacity\cite{10.3390/ijerph13050497} \\
\midrule
\textbf{Social} & Cognitive health &                                                                                                                                                                                                                                                                                                                                                                                                                                                                                                                                                                                                                                                                                                                                                                                                                                                                                                                                                                                                                                                                                                                                                                                                                                                                                                                                                                                  dementia prevention\cite{10.1002/gps.5626}, restorative effect against cognitive failures\cite{10.3390/ijerph15081705} \\
       & General health &                                                                                                                                                                                                                                                                                                                                                                                                                                                                                                                                                                                                                                                                                                                                                                                                                                                                                                                                                                                                                                                                                                                                                                                                                                                                                                                                                                                                                                                                  longevity\cite{10.1136/jech.56.12.913} \\
       & Mental health &                                                                                                                                                                                                                                                                                                                                                                                                                                                                                                                                                                                                                                                                                                                                                                                                                                                                                                                                                                                                                                                                                                                                                        mood improvement\cite{10.1140/epjds/s13688-021-00278-7}, depression prevention\cite{10.3390/ijerph182211746,10.1186/s12889-019-7171-9}, various\cite{10.3389/fpubh.2020.551453}, improved mental health inventory (mhi-5) scores\cite{10.1007/s00038-017-0963-8}, stress reduction\cite{10.3390/ijerph182211746,10.1016/j.scitotenv.2017.11.160} \\
       & Physical health &                                                                                                                                                                                                                                                                                                                                                                                                                                                                                                                                                                                                                                                                                                                                                                                                                                                                                                                                                                                                                                                                                                                                                                                                                                                               access to healthy produce\cite{10.1111/inm.13149}, various\cite{10.3389/fpubh.2020.551453}, inflammation reduction\cite{10.3390/ijerph15081705}, increase of physical activity\cite{10.1186/s12889-019-7416-7,10.1186/s12889-021-10177-1} \\
       & Social health &                                                                                                                                                                                                                                                                                                                                                                                                                                                                                                                                                                                                                                                                                                                                                                                                                                                                                                                                                       social cohesion\cite{10.1002/ajcp.12559,Jennings2019,10.1140/epjds/s13688-021-00278-7,10.1016/j.scitotenv.2017.11.160,10.1111/inm.13149,10.1016/j.healthplace.2016.08.011}, various\cite{10.1016/j.ufug.2020.126888}, access to healthy produce\cite{10.1111/inm.13149}, increased social capital\cite{10.1016/j.socscimed.2018.04.051}, social loneliness reduction\cite{10.1016/j.healthplace.2008.09.006}, improve sense of social belonging\cite{10.1007/s10393-014-0939-6,10.3389/fpubh.2022.1105473,10.1016/j.envres.2018.09.033} \\
       & Well-being &                                                                                                                                                                                                                                                                                                                                                                                                                                                                                                                                                                                                                                                                                                                                                                                                                                                                                                                                                                                                                                                                                                                                                                                                                                                                                      increase restorative capacity\cite{10.3390/ijerph13050497}, quality of life\cite{10.1002/gps.5626,10.1186/s12889-021-10177-1,10.1016/j.scs.2020.102236}, enhanced social interactions\cite{10.3390/ijerph19031466} \\
       \midrule
\textbf{Cultural} & Cognitive health &                                                                                                                                                                                                                                                                                                                                                                                                                                                                                                                                                                                                                                                                                                                                                                                                                                                                                                                                                                                                                                                                                                                                                                                                                                                                                                                                                                                                                                                              dementia prevention\cite{10.1002/gps.5626} \\
& General health &                                                                                                                                                                                                                                                                                                                                                                                                                                                                                                                                                                                                                                                                                                                                                                                                                                                                                                                                                                                                                                                                                                                                                                                                                                                                                                                                                                                                                                                                    various\cite{10.3390/ijerph13020196} \\
& Physical health &                                                                                                                                                                                                                                                                                                                                                                                                                                                                                                                                                                                                                                                                                                                                                                                                                                                                                                                                                                                                                                                                                                                                                                                                                                                                                                                                                                                                                                     increase of physical activity\cite{10.1016/j.scitotenv.2021.148293} \\
& Well-being &                                                                                                                                                                                                                                                                                                                                                                                                                                                                                                                                                                                                                                                                                                                                                                                                                                                                                                                                                                                                                                                                                                                                                                                                                                                                                                                                                                                                        quality of life\cite{10.1002/gps.5626,10.1016/j.scs.2020.102236,10.1016/j.scitotenv.2021.148293} \\
\midrule
\textbf{Mindfulness} & Mental health &                                                                                                                                                                                                                                                                                                                                                                                                                                                                                                                                                                                                                                                                                                                                                                                                                                                                                                                                                                                                                                                                                                                                                                                                                                                                                                            stress reduction\cite{10.3390/ijerph191711059}, anxiety reduction\cite{10.1016/j.envres.2022.114081}, depression prevention\cite{10.1080/15622975.2022.2112074,10.1016/j.envres.2022.114081} \\
& Physical health &                                                                                                                                                                                                                                                                                                                                                                                                                                                                                                                                                                                                                                                                                                                                                                                                                                                                                                                                                                                                                                                                                                                                                                                                                                                                                                                                                                                                                                           increase of physical activity\cite{10.1186/s12889-019-7416-7} \\
& Well-being &                                                                                                                                                                                                                                                                                                                                                                                                                                                                                                                                                                                                                                                                                                                                                                                                                                                                                                                                                                                                                                                                                                                                                                                                                                                                                                                                                                                                                                                     quality of life\cite{10.1016/j.explore.2012.12.002} \\
\bottomrule
\end{tabularx}

%% file: tables/LLM_benchmark.tex
\begin{tabular}{lcrrr}
  \toprule
LLM & Definitions & Temperature & $F_1$-score Main Activity & $F_1'$-score Weighted Combination \\ 
  \midrule
gpt-4 & \XSolidBrush & 0.9 & 0.772 & 0.772 \\ 
  gpt-4 & \XSolidBrush & 0.6 & 0.770 & 0.770 \\ 
  gpt-4 & \Checkmark & 0.3 & 0.764 & 0.764 \\ 
  gpt-4 & \XSolidBrush & 0.3 & 0.755 & 0.755 \\ 
  gpt-3.5-turbo & \XSolidBrush & 0.6 & 0.747 & 0.747 \\ 
  gpt-4 & \Checkmark & 0.6 & 0.740 & 0.740 \\ 
  gpt-3.5-turbo & \XSolidBrush & 0.9 & 0.728 & 0.728 \\ 
  gpt-3.5-turbo & \Checkmark & 0.3 & 0.726 & 0.726 \\ 
  gpt-4 & \Checkmark & 0.9 & 0.713 & 0.713 \\ 
  gpt-3.5-turbo & \Checkmark & 0.6 & 0.710 & 0.710 \\ 
  gpt-3.5-turbo & \XSolidBrush & 0.3 & 0.704 & 0.708 \\ 
  gpt-3.5-turbo & \Checkmark & 0.9 & 0.689 & 0.689 \\ 
   \bottomrule
\end{tabular}

%% file: tables/no_residual_correlations.tex
\begin{tabular}{lr}
  \toprule
\textbf{Activity Category} & \textbf{Mean Pearson Correlation} \\ 
  \midrule
  Physical & 0.16 \\ 
  Nature-appreciation & 0.05 \\ 
  Environmental & 0.18 \\ 
  Social & 0.06 \\ 
  Cultural & 0.18 \\ 
   \bottomrule
\end{tabular}